\documentclass[12pt,preprint]{aastex}
\usepackage{epsf}
\newcommand{\simgt}{\lower.5ex\hbox{$\; \buildrel > \over \sim \;$}}
\newcommand{\simlt}{\lower.5ex\hbox{$\; \buildrel < \over \sim \;$}}
\newcommand{\kperpbf}{\mathbf{k}_{\perp}}
\newcommand{\kappath}{\kappa_{\scriptscriptstyle \theta}}
\newcommand{\thetabf}{\mbox{\boldmath$\theta$}}
\newcommand{\thetath}{\theta_{\rm\scriptscriptstyle TH}}
\newcommand{\kperp}{k_{\perp}}

\newcommand{\kappamin}{\kappa_{\rm min}}
\newcommand{\kappamax}{\kappa_{\rm max}}
\newcommand{\kappaempty}{\kappa_{\scriptscriptstyle\rm empty}}
\newcommand{\var}{\langle\kappa^2\rangle}
\newcommand{\derivar}{\langle(\nabla\kappa)^2\rangle}
\newcommand{\sigmaln}{\sigma_{\rm ln}}
\newcommand{\zs}{z_{\rm s}}
\slugcomment{RESCEU-21/01\quad UTAP-402}
\shorttitle{lognormal property of weak lensing fields}
\shortauthors{Taruya, et al.}
\begin{document}
%
%
%
%
\title{\large Lognormal property of weak lensing fields}
\author{Atsushi Taruya}
\affil{ 
Research Center for the Early Universe (RESCEU), School of Science, \\
University of Tokyo, Tokyo 113-0033, Japan}
\email{ataruya@utap.phys.s.u-tokyo.ac.jp}
\author{Masahiro Takada\altaffilmark{1},~~ Takashi Hamana}
\affil{
National Astronomical Observatory, Mitaka, Tokyo 181-8588, Japan }
\email{mtakada@th.nao.ac.jp,~~hamana@yukawa.kyoto-u.ac.jp}
\author{Issha Kayo}
\affil{Department of Physics, School of Science, University of Tokyo,
Tokyo 113-0033, Japan}
\email{kayo@utap.phys.s.u-tokyo.ac.jp}
\and
\author{Toshifumi Futamase}
\affil{Astronomical Institute, Tohoku University, Sendai 980-8578, Japan}
\email{tof@astr.tohoku.ac.jp}
\altaffiltext{1}{Present address: 
Department of Physics \& Astronomy, University of Pennsylvania, 
PA 19104, USA; 
mtakada@hep.upenn.edu}
\begin{abstract}
The statistical property of the weak lensing fields is studied
quantitatively using the ray-tracing simulations. 
Motivated by the empirical lognormal 
model that characterizes the probability distribution function of the 
three-dimensional mass distribution excellently, 
we critically investigate the validity of lognormal model in the 
weak lensing statistics. Assuming that the 
convergence field, $\kappa$, is approximately described by the lognormal
distribution, we present analytic formulae of convergence for the
one-point probability distribution function (PDF) and the Minkowski
functionals.  The validity of lognormal models is checked in detail by
comparing those predictions with ray-tracing simulations in various cold
dark matter models.  We find that the one-point lognormal PDF can 
describe the non-Gaussian tails of convergence fields accurately up to
$\nu\sim10$, where $\nu$ is the level threshold given by
$\nu\equiv\kappa/\var^{1/2}$, although the systematic deviation from
lognormal prediction becomes manifest at higher source redshift and
larger smoothing scales.  The lognormal formulae for Minkowski
functionals also fit to the simulation results when 
the source redshift is low, $\zs=1$. Accuracy of the lognormal-fit 
remains good even at the small angular scales
$2$'$\simlt\theta\simlt4$', where the perturbation formulae by Edgeworth
expansion break down. On the other hand, the lognormal 
model enables us to predict the higher-order moments, i.e., skewness,
$S_{3,\kappa}$ and kurtosis, $S_{4,\kappa}$, and we thus discuss the 
consistency by comparing the predictions with the simulation
results. Since these statistics are very sensitive to the 
high(low)-convergence tails, the lognormal prediction does not provide a 
quantitative successful fit. 
We therefore conclude that 
the empirical lognormal model of the convergence field is safely 
applicable as a useful cosmological tool, as long as we are concerned with 
the non-Gaussianity of $\nu\simlt5$ for low-$\zs$ samples.  

\end{abstract}
\section{Introduction} 
\label{sec: intro}

Cosmic shear, coherent distortions in galaxy images caused by the
gravitational field of the intervening large-scale structure, has now
been recognized as a powerful cosmological tool (see Mellier 1999 and
Bartelmann \& Schneider 2001 for reviews).  Since the signal of cosmic
shear reflects the gravitational potential of the total mass
distribution, the cosmic shear can be a direct probe of the dark matter
distribution as well as a way to measure the cosmological parameters.
Recently, measurements of the cosmic shear have been independently
performed by several groups (Van Waerbeke et al. 2000a; Wittman et
al. 2000; Bacon, Refregier \& Ellis 2000; Kaiser, Wilson \& Luppino
2000; Maoli et al. 2001; Rhodes et al. 2001; van Waerbeke et al. 2001a).
The estimated shear variance from the different observational data set
quantitatively reconciles with each other, which is in good agreement
with theoretical predictions based on the cluster normalized cold dark
matter (CDM) model of structure formation.  Ongoing and future
wide-field surveys using high-resolution CCD camera will promise to
dramatically improve the signal-to-noise ratio of the weak lensing
signals. Hence, a detail understanding of the weak lensing
statistics will be  necessary to extract all the cosmological information
present in the data.

Of course, simplest but useful statistical measure of the weak lensing
field is popular two-point statistics (Blandford et al. 1991;
Miralda-Escude 1991; Kaiser 1992), however, a more clear discriminator
is needed to break the degeneracy in constraining the cosmological
parameters. In this respect, the non-Gaussian features of weak lensing field 
can be very sensitive to the density of universe and 
the higher-order statistics such as the 
skewness of local convergence field are expected to constrain the density 
parameter $\Omega_0$ and even to the cosmological constant $\lambda_0$ 
(Bernardeau, Van Waerbeke \& Mellier 1997; Jain \& Seljak 1997; 
Van Waerbeke et al. 2001b). 
On the other hand, topological analysis using 
the Minkowski functionals of the convergence field 
has been proposed by Sato, Takada, Jing \& Futamase (2001) 
in order to extract the non-Gaussian features 
(see also Matsubara \& Jain 2001). 
Unfortunately, their methodology heavily relies on the validity of the 
perturbation theory of structure formation and cannot be applied to the
weak lensing field at small angular scales $\theta\simlt 4$', where the 
underlying three-dimensional mass distribution is in the highly 
nonlinear regime. While the non-Gaussian signature of weak lensing fields 
becomes easily measurable on small scales, the non-Gaussian signal 
primarily reflects the nonlinear growth of 
the density fields, which strongly depends on the dark matter 
clustering properties. It is therefore desirable to investigate the 
statistical properties of three-dimensional mass density and 
explore the relation between mass distribution and weak lensing field.

Very recently, Kayo, Taruya \& Suto (2001) performed a detailed analysis
of the one- and two-point statistics of the three-dimensional mass
distribution using the high-resolution N-body simulations. They found
that the lognormal models of one- and two-point probability distribution
functions (PDF) can provide an excellent approximation to the nonlinear
mass distribution with Gaussian initial condition, irrespective of the
shape of the initial power spectrum.  The lognormal distribution has
been long known as an empirical prescription characterizing the dark
matter distribution and/or the observed galaxies (e.g., Hamilton 1985;
Bouchet et al. 1993; Coles \& Jones 1991; Kofman et al. 1994), however,
there exists no rigorous explanation for its physical
origin. Nevertheless, the lognormal model is now widely utilized in the
astrophysical context such as analytical modeling of dark halo biasing
and Ly-$\alpha$ forest (e.g., Mo \& White 1996; Taruya \& Suto 2000; Bi
\& Davidsen 1997).  In this regard, the result of Kayo et al. (2001) is
interesting and can be useful in quantifying the dark matter
distribution. Furthermore, their results indicate that 
the lognormal distribution can also describe the weak lensing field, 
since the weak lensing effect primarily reflects the clustering property 
of dark matter distributions. Then, crucial but fundamental questions 
arise as follows. How does the lognormal property emerge in the weak 
lensing field ?  Can the lognormal model provide an approximate and reliable 
prescription for the weak lensing statistics ?

In this paper, we quantitatively investigate those issues using the
ray-tracing simulations. We first consider how the statistical feature
of three-dimensional mass distribution is related to that of the local
convergence, i.e, the two-dimensional projected density field.  Assuming
that the convergence field is well-approximated by the lognormal
distribution, we derive analytic formulae for the one-point PDF and
the Minkowski functionals.  We then perform quantitative comparisons between
the lognormal models and the results of ray-tracing simulations.
Furthermore, we discuss the consistency of the lognormal model with the
higher-order statistics on which the previous works have mainly focused.

The plan of paper is as follows. In section \ref{sec: preliminaries}, 
we briefly describe the basic definitions of weak lenses. 
Then, we discuss the statistical relations between mass 
density field and convergence field in section \ref{sec: lognormal}. 
The analytic expressions for 
the one-point PDF and the Minkowski functionals are presented. 
In section \ref{sec: comparison},   
the detailed comparison between the lognormal predictions and the 
ray-tracing simulations is described. The consistency between 
lognormal model and previous study is discussed in section 
\ref{sec: moments}.  
Finally, section \ref{sec: summary} is devoted to conclusions and 
discussion. 

\section{Preliminaries}
\label{sec: preliminaries}

Inhomogeneous mass distribution of the large-scale structure deflects
the light ray trajectory emitted from an angular direction $\thetabf$ in
the source plane by a small angle $\delta\thetabf$ in the image
plane. The differences between the deflection angles of light rays
emitted from a galaxy thus induce a distortion of the galaxy image
characterized by the following symmetric matrix:
\begin{equation}
  \label{eq: matrix}
  \Phi_{i,j}\equiv\frac{\partial\delta\theta_i}{\partial \theta_j}
  =-2\int_0^{\chi_s}d\chi\,\,\frac{r(\chi)r(\chi_s-\chi)}{r(\chi_s)}\,\,
  \partial_i\partial_j\,\phi(\chi),\,\,\,\,\,\,\,
\,\,\,(i,j=1,2),
\end{equation}
where $\partial_{i}$ represents a derivative with respect to $\theta_i$,
and $\phi$ is the gravitational potential of the three-dimensional mass
density field. The variable $\chi_s$ means the quantity $\chi$ at the 
source redshift, and the quantities $\chi$ and $r(\chi)$ respectively 
denote the comoving parts of the radial and the angular-diameter distance: 
\begin{equation}
  \chi(z)=\int_{0}^{z} \frac{c\,\, dz'}{H(z')}
  \,\,;\,\,\,
 H(z)=H_{0}\sqrt{\Omega_0(1+z)^{3}+(1-\Omega_0-\lambda_0)(1+z)^{2}+
    \lambda_{0}},
  \label{eq: radial}  
\end{equation}
and 
\begin{equation}
  r(\chi) = \,\left\{
\begin{array}{lr}
\sin{(\sqrt{K}\,\chi)}/\sqrt{K} & (K>0) \\
\chi & (K=0) \\
\sinh{(\sqrt{-K}\,\chi)}/\sqrt{-K} & (K<0) \\
\end{array}
\right. 
\end{equation}
with the quantity $K$ being the spatial curvature of the universe,
$K=(H_0/c)^2(\Omega_0+\lambda_0-1)$.  The matrix $\Phi_{i,j}$ is usually
decomposed into the trace part, i.e, the convergence field $\kappath$
and the tidal shear components $\mbox{\boldmath$\gamma$}_{\theta}$
defined by
\begin{eqnarray}
  \label{eq: kappa}
  \kappath &=&-\frac{1}{2}\left(\Phi_{1,1}+\Phi_{2,2}\right), 
\\
  \label{eq: shear}
  \mbox{\boldmath$\gamma$}_{\theta}&=&
  -\frac{1}{2}\left(\Phi_{1,1}-\Phi_{2,2}\right)-i\,\Phi_{1,2}.
\end{eqnarray}

The shear field $\mbox{\boldmath$\gamma$}_{\theta}$ can be directly
estimated from the observed ellipticity of the galaxy images. The
convergence field $\kappath$ can be then reconstructed from the shear
map from the relations (\ref{eq: kappa}) and (\ref{eq: shear}) (e.g., 
Kaiser \& Squires 1993).  Since
we are interested in the {\em statistical} properties of the weak
lensing fields, one can safely employ the Born approximation, where the 
quantity $\partial_i\partial_j\phi(\chi)$ are computed along the 
unperturbed photon trajectory (Blandford et al. 1991; 
Miralda-Escude 1991; Kaiser 
1992).  From equations (\ref{eq: matrix}) and (\ref{eq: kappa}),
the convergence field along a line-of-sight is simply expressed as a
weighted projection of the mass density fluctuation field $\delta$ 
(e.g, Mellier 1999; Bartelmann \& Schneider 2001 for reviews): 
\begin{equation}
 \kappath(\thetabf)=\int_0^{\chi_s} d\chi\,w(\chi,\chi_s)\,
\delta[r(\chi)\thetabf,\chi],  
\label{eq: def_kappa}
\end{equation}
where the weight function $w(\chi,\chi_s)$ is 
\begin{equation}
 w(\chi,\chi_s)=\frac{3}{2}\left(\frac{H_0}{c}\right)^2
(1+z)\,\Omega_0\,\,
\frac{r(\chi)r(\chi_s-\chi)}{r(\chi_s)}.  
\end{equation}
Note that, even in the weak lensing limit, the density fluctuation
$\delta$ is not small but can become much larger than unity.

In statistical analysis of weak lensing fields, the smoothing filter is
practically used in order to reduce the noise due to the
intrinsic ellipticity of galaxies.  Throughout this paper, we adopt the
top-hat smoothing function. The variance of the local convergence can 
then be expressed as
\begin{equation}
\var= 
\int_0^{\chi_s}d\chi\,\left[w(\chi,\chi_s)\right]^2\,
\int\frac{d^2\kperpbf}{(2\pi)^2}\,P_{\rm\scriptscriptstyle mass}(\kperp)\,\,
\hat{W}^2_{\rm TH}\left(\kperp\,r(\chi)\thetath\right),
\label{eq: var_kappa} 
\end{equation}
where $P_{\rm\scriptscriptstyle mass}(k)$ is the three-dimensional power
spectrum of the mass distribution and the function $\hat{W}_{\rm TH}(x)$
represents the Fourier transform of the top-hat smoothing kernel: 
\begin{equation}
 \hat{W}_{\rm TH}(x) =2\,\frac{J_1(x)}{x}.
\end{equation}
Note that equation (\ref{eq: var_kappa}) is derived by employing the
Limber's equation (Kaiser 1992).  Similarly, variance of the gradient
field, $\langle(\nabla\kappa)^2\rangle$, can be expressed as
\begin{equation}
\derivar = 
\int_0^{\chi_s}d\chi\,\left[w(\chi,\chi_s)\right]^2\,r^2(\chi)\,\,
\int\frac{d^2\kperpbf}{(2\pi)^2}\,\kperp^2 \,\,
P_{\rm\scriptscriptstyle mass}(\kperp)\,\,
\hat{W}_{\rm TH}^2\left(\kperp\,r(\chi)\thetath\right). 
\label{eq: var_nablakappa}
\end{equation}
Equations (\ref{eq: var_kappa}) and (\ref{eq: var_nablakappa}) are valid
on relevant angular scales we are interested in, where the small-angle
approximation holds with good accuracy (Hu 2001).

\section{Lognormal model prescription}
\label{sec: lognormal}

\subsection{One-point PDF}
\label{subsec: one-point}

The expression (\ref{eq: def_kappa}) gives a simple interpretation that
the statistical feature of $\kappa$ is closely related to that of the
underlying density field $\delta$.  To show this more explicitly, we
consider the one-point PDF, which can be constructed from a full set of
the moments.  Let us write down the PDF of the mass density field
$\delta$:
  \begin{equation}
    \label{eq: pdf_delta}
P(\delta)d\delta=\int_{-\infty}^{+\infty} 
\frac{dx}{2\pi \sigma^2}\,\,e^{-ix(\delta/\sigma)+
\varphi_{\delta}(x)/\sigma^2}\,\,d\delta, 
  \end{equation}
where $\sigma$ denotes the root-mean-square (RMS) of $\delta$, 
$\sigma=\langle\delta^{2}\rangle^{1/2}$ and  
$\varphi_{\delta}(x)$ means cumulant generating function:   
\begin{equation}
  \label{eq: cumulant_del}
  \varphi_{\delta}(x)=\sum_{n=2}^{\infty}\,
  \frac{S_{n,\delta}}{n!}\,(ix)^n
  \,\,;\,\,\,\,\,S_{n,\delta}\equiv\frac{\langle\delta
    ^n\rangle_{\rm c}}  {\langle\delta^2\rangle^{n-1}}.
\end{equation}
Similarly, the one-point PDF of the local convergence is 
written as 
  \begin{equation}
    \label{eq: pdf_kappa}
P(\hat{\kappa})d\hat{\kappa}=\int_{-\infty}^{+\infty}  
\frac{dx}{2\pi \langle\hat{\kappa}^2\rangle}\,\,
e^{-ix(\hat{\kappa}/\langle\hat{\kappa}^2\rangle^{1/2})+
\varphi_{\hat{\kappa}}(x)/\langle\hat{\kappa}^2\rangle}\,\,d\hat{\kappa}, 
  \end{equation}
with the cumulant generating function for $\hat{\kappa}$,
$\varphi_{\hat{\kappa}}(x)$.  Here we have introduced the normalized
convergence field, $\hat{\kappa}\equiv\kappa/|\kappamin|$ so as to
satisfy the range of the definition being $-1<\hat{\kappa}<+\infty$. The
quantity $\kappamin$ denotes the minimum value of the convergence
field. Note that the actual value of $\kappamin$ in the universe should,
in principle, depend on the nature of dark matter between the source
galaxies and observer (Metcalf \& Silk 1999; Seljak \& Holtz 1999).  If
the dark matter is composed of a compact object such as the primordial
black holes, $\kappamin$ might correspond to the convergence evaluated
along an empty beam (see eq.[\ref{eq: empty}]).  On the other hand, when
the dark matter is composed of a smooth microscopic component as is
suggested by the standard CDM scenarios, $\kappamin$ becomes larger than
the empty-beam value for relevant smoothing scales (see also Jain,
Seljak \& White 1999).  As will be discussed later in section
\ref{subsec: model_param}, $\kappamin$ is an important quantity in
controlling the lognormal model and is sensitive to the cosmological
model.

Even though we have the simple relation (\ref{eq: def_kappa}) 
between $\delta$ and $\kappa$, 
the relation between the  cumulant generating functions 
of $\varphi_{\delta}$ and $\varphi_{\hat{\kappa}}$ is practically 
intractable without recourse to some assumptions or approximations. 
Under the hierarchical ansatz for the higher order moments of $\delta$, i.e., 
$S_{n,\delta}=\mbox{const.}$, Munshi \& Jain (2000) and Valageas (2000) 
showed that the relation between $\varphi_{\delta}$ and 
$\varphi_{\hat{\kappa}}$ is greatly reduced and is expressed in a 
compact form:  
  \begin{equation}
    \label{eq: cumulant}
    \varphi_{\hat{\kappa}}(x)=\int_{0}^{\chi_s}d\chi\,
    \frac{\langle\hat{\kappa}^2\rangle}{I_{\kappa}(\theta)} \,\,
    \varphi_{\delta}\left(\frac{w(\chi,\chi_s)}{|\kappamin|}\,
      \frac{I_{\kappa}(\theta)}{\langle\hat{\kappa}^2\rangle}\,x\right), 
  \end{equation}
where the quantity $I_{\kappa}$ is defined by 
\begin{equation}
  \label{eq: I_kappa}
  I_{\kappa}(\theta)= 
\int\frac{d^2\kperpbf}{(2\pi)^2}\,P_{\rm\scriptscriptstyle mass}(\kperp)\,\,
\hat{W}^2_{\rm TH}\left(\kperp\,r(\chi)\theta\right).
\end{equation}
This is valid as long as the small-angle approximation holds. As
also proposed by those authors, since the
factor $w\,I_{\kappa}/(|\kappamin|\langle\hat{\kappa}^2\rangle)$ in
equation (\ref{eq: cumulant}) has typical values of order unity, one
might use a simple approximation 
%
\begin{equation}
  \label{eq: cumulant_approx}
  \varphi_{\hat{\kappa}}(x) \simeq \varphi_{\delta}(x). 
\end{equation}
%
It is easy to check that this approximation holds for the limit of
$z_{\rm s}\rightarrow 0$, where the both quantities $
w\,I_{\kappa}/(|\kappamin|\langle\hat{\kappa}^2\rangle)$ and $\int
\!d\chi\langle\hat{\kappa}^2\rangle/I_{\kappa}(\theta)$ approach unity.

Once equation (\ref{eq: cumulant_approx}) is given, PDF of the
convergence field, $P(\kappa)$, can be {\it directly} calculated from
the one-point PDF of the three-dimensional mass density field
$P(\delta)$, irrespective of the projection effect (see eq.[\ref{eq:
def_kappa}]):
\begin{equation}
  \label{eq: pdf_kappa_del}
  P(\kappa)\,\,d\kappa =\,P
  \left(\delta\to\frac{\kappa}{|\kappamin|};\,\,
    \sigma\to\frac{\var^{1/2}}{|\kappamin|}\right)
\,\,\frac{d\kappa}{|\kappamin|}  
\end{equation}
Now, let us recall the empirical fact that the PDF of local density 
$P(\delta)$ is approximately described by the lognormal distribution: 
\begin{equation}
  \label{eq: log_delta}
P_{\rm ln}(\delta)\,\,d\delta=
\frac{1}{\sqrt{2\pi\ln(1+\sigma^2)}}\,\,\exp
\left\{-\frac{[\ln(1+\delta)\sqrt{1+\sigma^2}]^2}{2\ln(1+\sigma^{2})}
\right\}\,\,\frac{d\delta}{1+\delta}.
\end{equation}
Substituting (\ref{eq: log_delta}) into (\ref{eq: pdf_kappa_del}), one
finally obtains
\begin{equation}
  \label{eq: logpdf}
P_{\rm ln}(\kappa)\,\,d\kappa=
\frac{1}{\sqrt{2\pi}\sigmaln}\,\,\exp
\left\{-\frac{[\ln(1+\kappa/|\kappamin|)+\sigmaln^2/2]^2}{2\sigmaln^2}
\right\}\,\frac{d\kappa}{\kappa+|\kappamin|}, 
\end{equation}
where the quantity $\sigmaln$ is defined by 
\begin{equation}
  \sigmaln^{2}\equiv \ln\left(1+\frac{\var}{|\kappamin|^{2}}\right) . 
\end{equation}

Before discussing the application and validity of equation (\ref{eq:
logpdf}), two important remarks should be mentioned. First, recall that
the lognormal distribution violates the assumption of hierarchical
ansatz used in the derivation of equation (\ref{eq: cumulant}). 
The resultant expression (\ref{eq: logpdf}) is thus inconsistent, 
however, the violation of the hierarchical assumption is fortunately 
weak on nonlinear scales of interest here (e.g., Kayo et al. 2001).  We
therefore expect that, as long as we restrict ourselves to the range of
applicability, the expression (\ref{eq: logpdf}) can provide a
reasonable approximation and capture an important aspect of the
non-Gaussianity in the weak lensing field.  Secondly, notice that the
result (\ref{eq: logpdf}) heavily relies on the validity of the
approximation (\ref{eq: cumulant_approx}). One might suspect that the
approximation breaks down for a case with the high source redshift such
as $\zs\simgt 1$, where the lensing projection becomes more important,
and this will be discussed below.  Of course, one can directly evaluate
the convergence PDF $P(\kappa)$ from equations (\ref{eq: pdf_kappa}) and
(\ref{eq: cumulant}) assuming the lognormal PDF of $\delta$, although
this treatment is not useful in practice. Instead, in this paper 
we {\em a priori} assumes the simple analytic prediction 
(\ref{eq: logpdf}), whereby we can further derive the useful 
analytic formulae
for the Minkowski functionals and obtain a more physical interpretation 
of the results. We will then carefully check our model by comparing the
predictions with the numerical simulation results.

\subsection{Minkowski functionals}
\label{subsec: minkowski}

Expression (\ref{eq: pdf_kappa_del}) implies that the projection effect
is primarily unimportant and is eliminated by rescaling the quantities
$\delta\to\kappa/|\kappamin|$. That is, the statistical feature of
$\kappa$ directly reflects that of three-dimensional density $\delta$.
If this is the case, one can further develop a prediction of the weak
lensing field based on the lognormal ansatz. To investigate this issue,
the other informative statistics such as the higher-order correlation
and the isodensity statistics should be examined.

Among these statistics, it is known that Minkowski functionals can be
useful and give a morphological description to the contour map of the
weak lensing field (e.g, Schmalzing \& Buchert 1997). In a
two-dimensional case, Minkowski functionals are characterized by the
three statistical quantities: the area fraction, $v_0$, circumference
per unit length, $v_1$ and Euler characteristics per unit area, $v_2$.
The third functional is equivalent to the famous genus statistics often used
in the cosmological context (Gott, Melott \& Dickinson 1986).  These
quantities are evaluated for each isocontour as a function of level
threshold, $\nu\equiv\kappa/\langle\kappa^2\rangle$.

The explicit expression of Minkowski functional $v_0(\nu)$ 
  is given by 
  \begin{equation}
    \label{eq: def_v0}
    v_0(\nu) 
    = \left\langle\Theta\left(\kappa-\nu\,\langle\kappa^2\rangle^{1/2}\right)
    \right\rangle, 
  \end{equation}
  where $\Theta$ denotes the Heaviside step function. 
  The Minkowski functionals $v_1(\nu)$ and $v_2(\nu)$ provide an 
  additional information on the convergence field, since the definitions
  include higher derivative terms such as $\partial_i\kappa$ and
  $\partial_{i}\partial_{j}\kappa$.  According to Matsubara (2000),
  Minkowski functionals $v_1$ and $v_2$ in the two-dimension map can be
  related to the level-crossing and the genus statistics defined in the
  two-dimensional surface, $N_1$ and $G_2$ (see also Schmalzing \&
  Buchert 1997):
\begin{eqnarray}
  v_{1}(\nu) &=& \frac{\pi}{8}\,\,N_{1}(\nu) 
  = \frac{\pi}{8}\left\langle \delta_{D}\left
      (\kappa-\,\langle\kappa^2\rangle^{1/2}\nu\right)\,\,
  |\partial_1\kappa|\right\rangle,
\label{eq: def_v1} \\
  v_{2}(\nu) &=& G_{2}(\nu)= \left\langle
\delta_{D}\left
      (\kappa-\,\langle\kappa^2\rangle^{1/2}\nu\right)\,\,
    \delta_D(\partial_1\kappa)\,\,|\partial_{2}\kappa|\,\,
    \partial_{1}\partial_{1}\kappa
\right\rangle.
\label{eq: def_v2}
\end{eqnarray}

Assuming that the convergence field is approximately described by the
lognormal distribution, the analytic expressions for Minkowski
functionals can be derived in a straightforward manner.  Since the area
fraction represents the cumulative probability above the threshold
$\nu$, substituting (\ref{eq: logpdf}) into the definition (\ref{eq:
def_v0}) yields
\begin{equation}
  v_{0,{\rm ln}}(\nu) 
 = \int_{\nu\,\langle\kappa^{2}\rangle^{1/2}}^{+\infty}\,d\kappa
    \,\,P_{\rm ln}(\kappa)
=\frac{1}{2}\,\mbox{erfc}\,\left\{\frac{y(\nu)}{\sqrt{2}}\right\}
  \label{eq: v0_log}
\end{equation}
with the function $y(\nu)$ given by
\begin{equation}
  \label{eq: def_y}
y(\nu)\equiv \frac{\sigmaln}{2}+\frac{\ln(1+\nu\,\var^{1/2}/|\kappamin|)}
{\sigmaln}.
\end{equation}
As for the Minkowski functionals $v_1$ and $v_2$, the analytic
expressions for the lognormal distribution can be obtained from the
local transformation of Gaussian formulae, $N_{1}(\nu)$ and
$G_{2}(\nu)$.  The details of the derivation are described by Taruya \&
Yamamoto (2001) (see also Matsubara \& Yokoyama 1996).  The final
results become
\begin{eqnarray}
  \label{eq: v1_log}
v_{1,{\rm ln}}(\nu) &=& \frac{1}{8\sqrt{2}}\,\, 
\frac{1}{\sqrt{|\kappa_{\rm min}|^2+\var}}\,
\frac{\derivar^{1/2}}{\sigmaln}\,\,\,\,e^{-y^2(\nu)/2}, 
\\
  \label{eq: v2_log}
v_{2,{\rm ln}}(\nu) &=& 
\frac{1}{2(2\pi)^{3/2}} \,\,\frac{1}{|\kappa_{\rm min}|^2+\var}\,\,
\frac{\derivar}{\sigmaln^2}\,\,y(\nu)\,\,e^{-y^2(\nu)/2}. 
\end{eqnarray}
Note that in the limit $\langle\kappa^{2}\rangle^{1/2}/|\kappamin|\ll1$,
the above predictions recover the Gaussian formulae:
\begin{eqnarray}
  \label{eq: v0_g}
  v_{0,{\scriptscriptstyle\rm G}}(\nu) &=&
\frac{1}{2}\,\, \mbox{erfc}\left(\frac{\nu}{2}\right), 
\\
  \label{eq: v1_g}
  v_{1,{\scriptscriptstyle \rm G}}(\nu) &=&
\frac{1}{8\sqrt{2}} \,\,
  \left(\frac{\derivar}{\var}\right)^{1/2}\,\,e^{-\nu^2/2}, 
\\
  \label{eq: v2_g}
  v_{2,{\scriptscriptstyle \rm G}}(\nu) &=&
\frac{1}{2(2\pi)^{3/2}} \,\,
  \frac{\derivar}{\var}\,\,\nu\,e^{-\nu^2/2}. 
\end{eqnarray}

Apart from the PDF $P(\kappa)$ and the area fraction $v_0$, there exists
no clear reason that the extension of lognormal model to the Minkowski
functionals $v_1$ and $v_2$ still provides a good approximation in the
real universe. At present, the predictions (\ref{eq: v1_log}) and
(\ref{eq: v2_log}) are just regarded as an {\it extrapolation} from the
one-point PDF $P(\kappa)$ and should be checked by numerical
simulations.

\section{Comparison with ray-tracing simulations}
\label{sec: comparison}

In this section, we quantitatively examine the validity of lognormal
model by comparing its predictions with the simulation results.  A brief
description of ray-tracing simulation is presented in section
\ref{subsec: simulation}.  In section \ref{subsec: model_param}, 
the model parameters in the lognormal formulae are
checked in details using the simulation data. Some important numerical
effects are also discussed. Then, section \ref{subsec: results}
describes our main results, i.e, the comparisons of one-point PDF and
Minkowski functionals between the lognormal predictions and the
simulation data.

\subsection{Ray-tracing simulations}
\label{subsec: simulation}

In order to investigate the lognormal property of convergence fields, 
we use a series of ray-tracing simulations in three cold dark matter
models (SCDM, LCDM, OCDM for Standard, Lambda, Open CDM models, respectively). 
The cosmological parameters used here are summarized in Table 
\ref{tbl : parameter}. 

To perform a ray-tracing simulation, a light-cone data set is first
generated by particle-mesh (PM) N-body code. The PM code uses
$256^2\times 512$ particles and is performed in a periodic rectangular
box of size $(L,L,2L)$ with the force mesh $256^2\times 512$. The
initial conditions are generated according to the transfer functions of
Bond \& Efstathiou (1984). Then the light-cone of the particles is
extracted from each simulation during the run (Hamana, Colombi \& Suto
2001). The final set of light-cone data are created so as to cover a
field of view of $5\times5$ square degrees and the box sizes of each
output are chosen so as to match to the convergence of the light ray
bundle.  As a result, the angular resolution of ray-tracing simulation,
which is basically limited by the spatial resolution of PM N-body code,
becomes almost constant, $\theta_{\rm res}\approx 1.5$' from the
observer at $z=0$ (Hamana \& Mellier 2001).

Once the light-cone data is obtained, ray-tracing simulations are next
performed using the multiple lens-plane algorithm (e.g., Schneider,
Ehlers \& Falco 1992; Jain, Seljak \& White 2000; Hamana, Martel \&
Futamase 2000).  In our calculation, the interval between lens planes
are fixed to comoving length, $80 h^{-1}$Mpc. The $512^{2}$ light rays
are then traced backward from the observer's point to the source
plane. The initial ray directions are set on $512^{2}$ grids and the
two-dimensional deflection potential are calculated by solving
ray-bundle equations keeping the same grids in each lens plane.  We
obtained $40$ realizations by randomly shifting the simulation boxes.

After constructing the weak lensing map in the image plane at $z=0$, the
smoothed convergence field are finally computed on $512^{2}$ grids by
convolving the top-hat smoothing kernel. All the statistical quantities such
as one-point PDF and Minkowski functionals are evaluated from the data.
In a subsequent analysis, we use the convergence data set fixing the
source redshifts to $\zs=1$ and $2$.

\subsection{On the lognormal model parameters} 
\label{subsec: model_param}

Since the lognormal predictions presented in previous section heavily
rely on the three parameters, $\kappamin$, $\var$ and $\derivar$, we
first check them in some details using the ray-tracing simulations.

\begin{figure}[tbp]
  \begin{center}
   \leavevmode
\epsfxsize=11.0cm 
\epsfbox{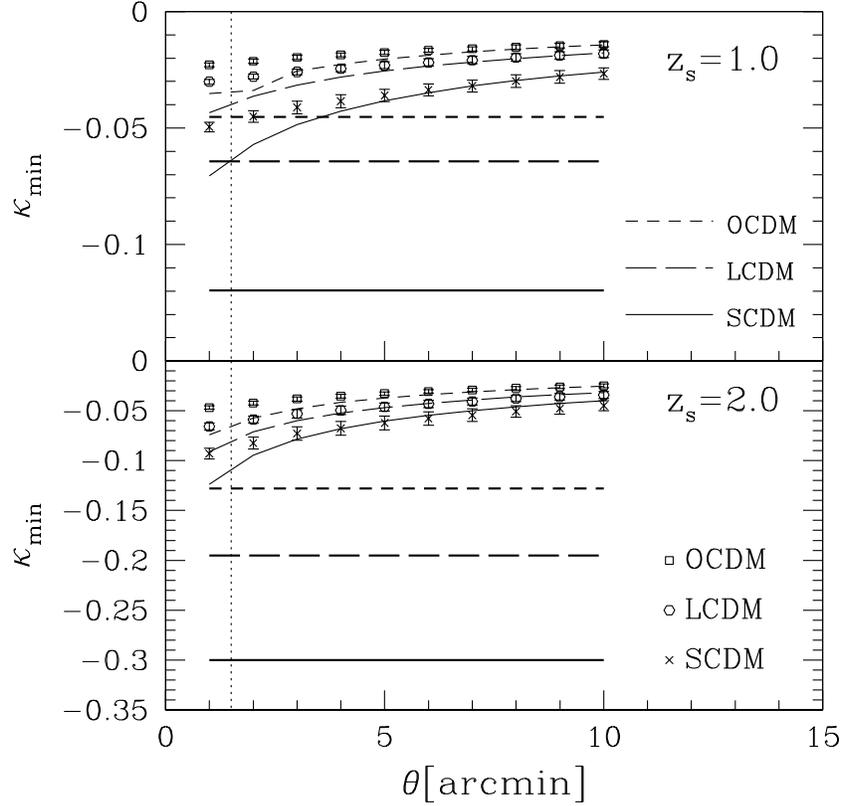}
    \caption{ Minimum values of the convergence field, $\kappamin$, 
      evaluated from the simulations. Thick lines indicate the 
      predictions assuming the empty beam (see eq.[\ref{eq: empty}]).  
      The thin lines represent the estimation taking account of the 
      effect of finite sampling (see Appendix A). The vertical dotted lines 
      denote the resolution limit of ray-tracing simulation, which 
      is primarily determined by the force resolution of PM N-body code 
      (see text in Sec.\ref{subsec: simulation}): 
      $\zs=1.0$ ({\it upper panel}); $\zs=2.0$ ({\it lower panel}). 
      \label{fig: kmin}}
  \end{center}
\end{figure}

Figure \ref{fig: kmin} shows the minimum value of the convergence,
$\kappamin$ against the smoothing angle. The error-bars indicate the
$1$-$\sigma$ errors around each mean, where the mean value $\kappamin$
is obtained from $40$ realization data. As stated in section
\ref{subsec: one-point}, the convergence $\kappa$ evaluated along an 
empty beam is theoretically considered as a possibility of $\kappamin$ and
expressed from equation (\ref{eq: def_kappa}) as
\begin{equation}
  \kappaempty
  = -\int_0^{\chi_s} d\chi\,w(\chi,\chi_s),
\label{eq: empty}  
\end{equation}
which implies that the light ray propagates through empty space with 
$\delta=-1$ everywhere along the line of sight.  In 
figure \ref{fig: kmin}, thick lines represent the theoretical prediction
(\ref{eq: empty}) for each model.  Clearly, these
predictions give much smaller values than the simulation results in 
both $\zs=1.0$ and $2.0$ cases. In addition to the systematic cosmological 
model dependences, the scale-dependence of $\kappamin$ also appears.

These results mean that none of the light rays becomes 
completely empty beam (see also Jain, Seljak \& White 1999). 
Nevertheless, one cannot exclude the possibility that some or even the 
majority of lines of sights become empty at scales smaller than
the mean separation angle between particles of N-body simulations. 
In this sense, the minimum value $\kappamin$ could be affected by the
finite sampling from a limiting survey size of the convergence map. 
Based on this consideration, we propose the following intuitive way to
explain the behaviors of $\kappamin$ shown in figure \ref{fig: kmin}.   
Assuming that the original PDF of {\em infinitesimal} light rays obeys 
the lognormal model with the minimum value $\kappaempty$,  
one can roughly estimate $\kappamin$ from the condition
that the expectation number of independent sampling area for
$\kappaempty\le\kappa\le\kappamin$ becomes unity in the field-of-view 
$5^{\circ}\times5^{\circ}$. 
The details of estimation is described in Appendix A. For a given angular 
scale $\thetath$ and cosmological model, the quantity $\kappamin$ is 
evaluated by solving the equation (\ref{eq: kmin}) with the prior PDF 
(\ref{eq: prior_pdf}). 
The thin lines in figure \ref{fig: kmin} depict the estimation based on this  
prescription, which reasonably agrees with simulations. This result 
is interesting in the sense that we could analytically predict 
$\kappamin$ and the resultant minimum value is sensitive to the 
cosmological parameters, especially $\Omega_0$. The practical 
possibility will be again discussed in the final section.

Next, turn to focus on the variances of the convergence and its gradient
fields.  In figure \ref{fig: varkappa}, the measured amplitudes,
$\var^{1/2}$ and $\derivar^{1/2}$, are plotted in linear- and log-scale
and are compared with linear ({\it short-dashed}) and nonlinear ({\it
solid}) predictions. For the RMS of the 
convergence $\var^{1/2}$, the nonlinear prediction based on the Peacock
\& Dodds (1996) formula faithfully reproduces the simulation result over
the smoothing angles, $\theta\simgt\theta_{\rm res}\approx 1.5$'.  On
the other hand, in the case of $\derivar$, nonlinear predictions
systematically deviate from simulations. The discrepancy remains even at
larger smoothing angles, $\theta\sim 10$' and causes $30\sim40\%$ error.

While the theoretical predictions based on the expressions (\ref{eq:
var_kappa}) and (\ref{eq: var_nablakappa}) were computed assuming that
the mass power spectrum $P_{\rm mass}(k)$ are continuous and has
infinite resolution, the simulation data is practically affected by the
finite resolution.  In our case, the limitation of PM force mesh could
be attributed to the finite resolution or the cutoff of Fourier modes,
which becomes influential even on larger smoothing angles,
$\theta\simgt\theta_{\rm res}$.  
In fact, compared to $\var$, dominant contribution to the quantity 
$\derivar$ comes from the short-wavelength modes of three-dimensional 
density fluctuations, sensitively depending on the choice of the smoothing 
filter. To show the significance of this effect, nonlinear prediction 
is modified according to the finite resolution of PM code (see Appendix B). 
Shortly, the Fourier-integrals in the expressions of 
(\ref{eq: var_kappa}) and (\ref{eq: var_nablakappa}) are discretized so as to 
mimic the PM N-body resolution as follows 
(Eqs.[\ref{eq: discretize}][\ref{eq: Delta}] in Appendix B): 
\begin{eqnarray}
&& \int_{0}^{\infty} d\kperp\,\,\,\Delta^{2}(\kperp)
W_{\rm TH}^{2}(\kperp r \theta) 
\,\,\,\,\, \Longrightarrow\,\,\,\,\,
\sum_{n=1}^{N_{\rm max}} \Delta \kperp\,\,\,\Delta^{2}
(k_{\perp,n})\,W_{\rm TH}^{2}(k_{\perp,n}\,r\theta),
  \nonumber 
\\
&&  \Delta^{2}(\kperp) \equiv\left\{
\begin{array}{cl}
(\kperp/2\pi)\,\,P_{\rm\scriptscriptstyle mass}(\kperp)
& \,\,\,\,\,\,\,\,\,\,\,\,\mbox{for} \,\,\,\var,
\\
\\
(\kperp^3/2\pi)\,\,P_{\rm\scriptscriptstyle mass}(\kperp)
& \,\,\,\,\,\,\,\,\,\,\,\,\mbox{for} \,\,\,\derivar.
\end{array}
\right.,  
  \nonumber
\end{eqnarray}
where $n$-th Fourier mode $k_{\perp,n}$ is  given by 
$k_{\perp,n}=\Delta\kperp\times n$ and the interval $\Delta\kperp$ is set to 
$\Delta\kperp=2\pi/L_{\rm box}(z)$.

The long-dashed lines in figure \ref{fig: varkappa} show the results 
including the finite-resolution effect. 
In each panel, the number of Fourier modes, $N_{\rm max}$ is chosen as $90$ 
(see discussion in Appendix B). It is apparent that the 
prediction of $\derivar^{1/2}$ systematically reduces its power all over
the scales and it almost reconciles with the simulation result, although
the result of $\var^{1/2}$ remains unchanged. We have also examined various 
cases by changing the maximum number to $64\simlt N_{\rm max}\simlt 128$, but 
obtained qualitatively similar behavior. 
Note that the incorrect prediction of 
$\derivar$ leads to a systematic error in predicting the amplitude of the 
Minkowski functionals 
$v_1$ and $v_2$ from their definitions (\ref{eq: v1_log}) and (\ref{eq: v2_log}), 
while it does not alter the non-Gaussian shapes of those functionals with 
respect to the level threshold.

Keeping up the above remarks in mind, in what follows, to make a
comparison with the lognormal models transparent, we use the
parameters $\var$, $\derivar$ and $\kappamin$ directly estimated from
  the simulations when plotting the lognormal predictions (\ref{eq:
  logpdf}), (\ref{eq: v0_log}), (\ref{eq: v1_log}) and (\ref{eq:
  v2_log}).

\subsection{Results} 
\label{subsec: results}

\subsubsection{One-point PDF} 
\label{subsec: one-point_sim}

\begin{figure}
  \begin{center}
    \leavevmode
    \epsfxsize=15.5cm 
    \epsfbox{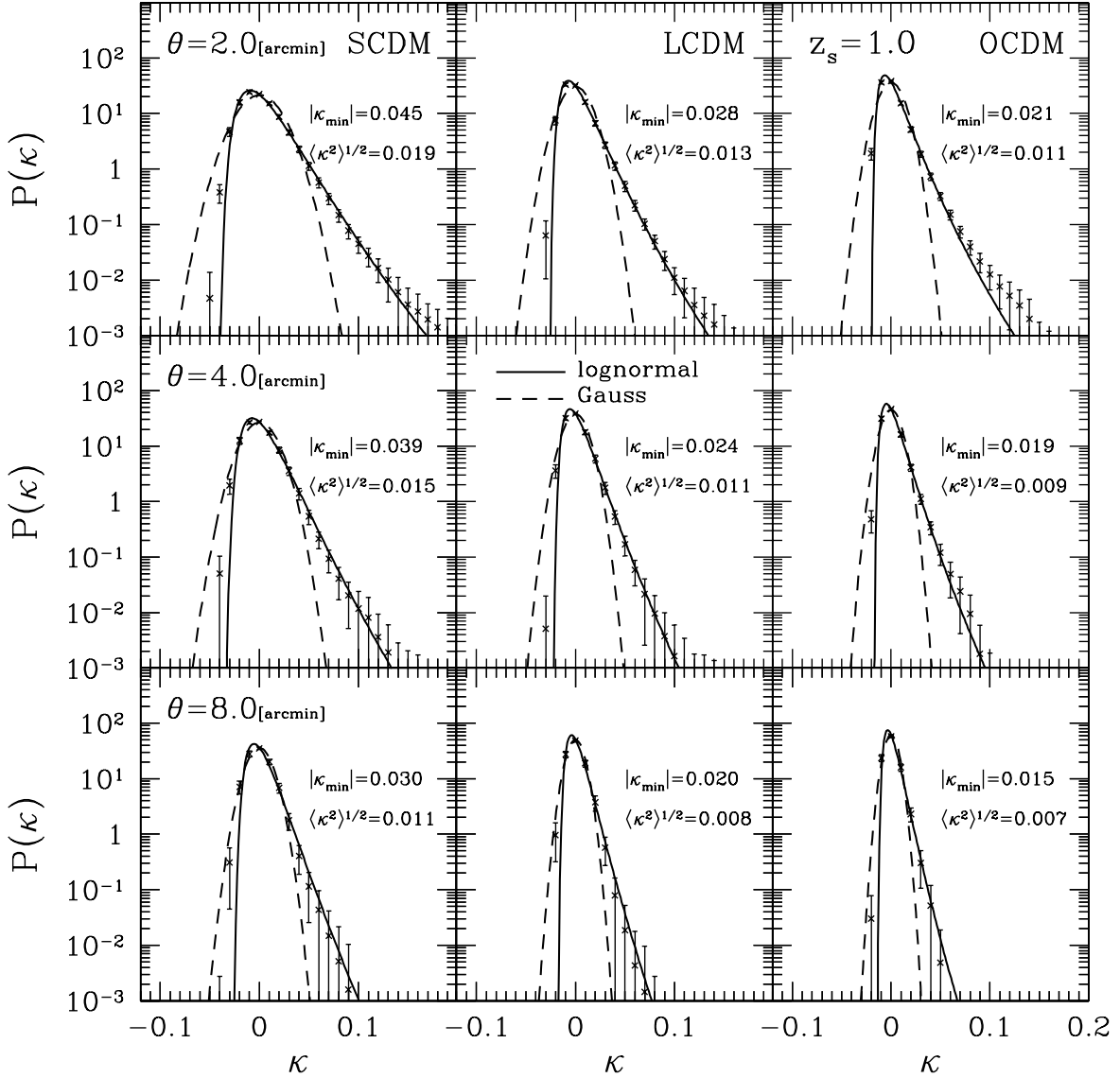}
  \end{center}
  \caption{One-point PDFs of the convergence field fixing the source redshift 
    to $\zs=1.0$ with smoothing angle $\theta=2$', $4$'
    and $8$'({\it top to bottom}). Solid lines show the lognormal 
    prediction (eq.[\ref{eq: logpdf}]),  
    where the parameters $\kappa$ and $\var$ are directly estimated from 
    simulations. For comparison, the Gaussian PDF with the same 
    variance $\var$ are plotted as dashed lines: 
    SCDM model ({\it left}) ; 
    LCDM model ({\it middle}) ; OCDM model ({\it right}). 
    \label{fig: pdf_z1} }
\end{figure}

\begin{figure}
  \begin{center}
    \leavevmode
    \epsfxsize=15.5cm 
    \epsfbox{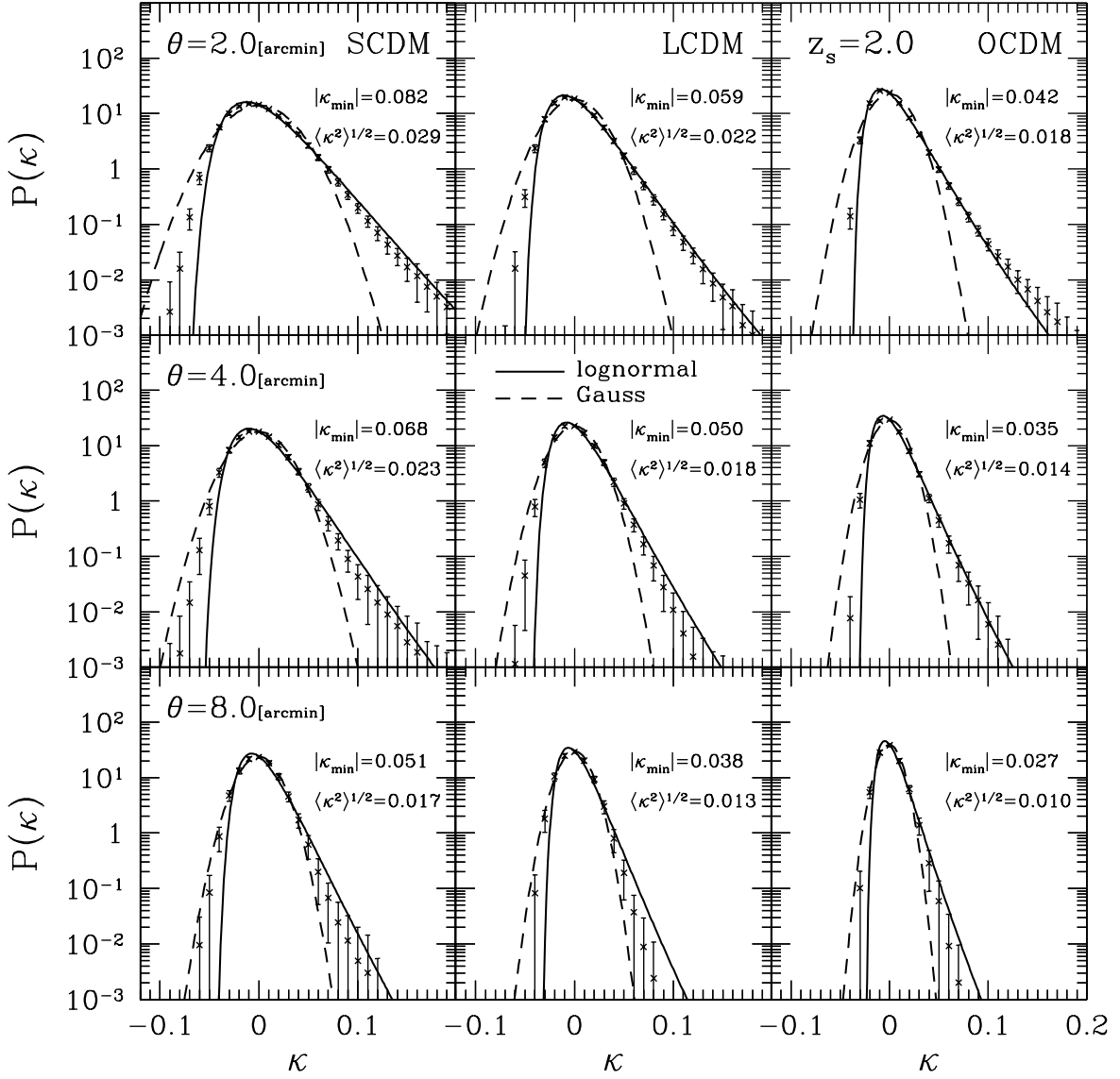}
  \end{center}
  \caption{Same as Figure \ref{fig: pdf_z1}, but we fix 
    the source redshift $\zs=2.0$.\label{fig: pdf_z2} }
\end{figure}

As a quick view of the validity of lognormal model, we first deal with
the one-point PDF, $P(\kappa)$.  Figures \ref{fig: pdf_z1} and \ref{fig:
pdf_z2} show the one-point PDFs of the local convergence in various 
CDM models with the smoothing angles, $\theta=2$', $4$' and $8$'. 
Here, the PDF data is constructed by binning the data with 
$\Delta\kappa=0.01$. 
The source redshifts are fixed to $\zs=1.0$ in 
figure \ref{fig: pdf_z1} and $\zs=2.0$ in figure \ref{fig: pdf_z2}. 
Clearly, the functional form of the 
one-point PDF becomes broader as increasing the source redshift
$\zs$. On small angular scales, the PDF significantly deviates from the
Gaussian PDF denoted by dashed lines. Although local convergences on
larger smoothing scales tend to approach the Gaussian form, they still
exhibit a non-Gaussian tail.

\begin{figure}
  \begin{center}
    \noindent
    \epsfxsize=10.5cm 
    \epsfbox{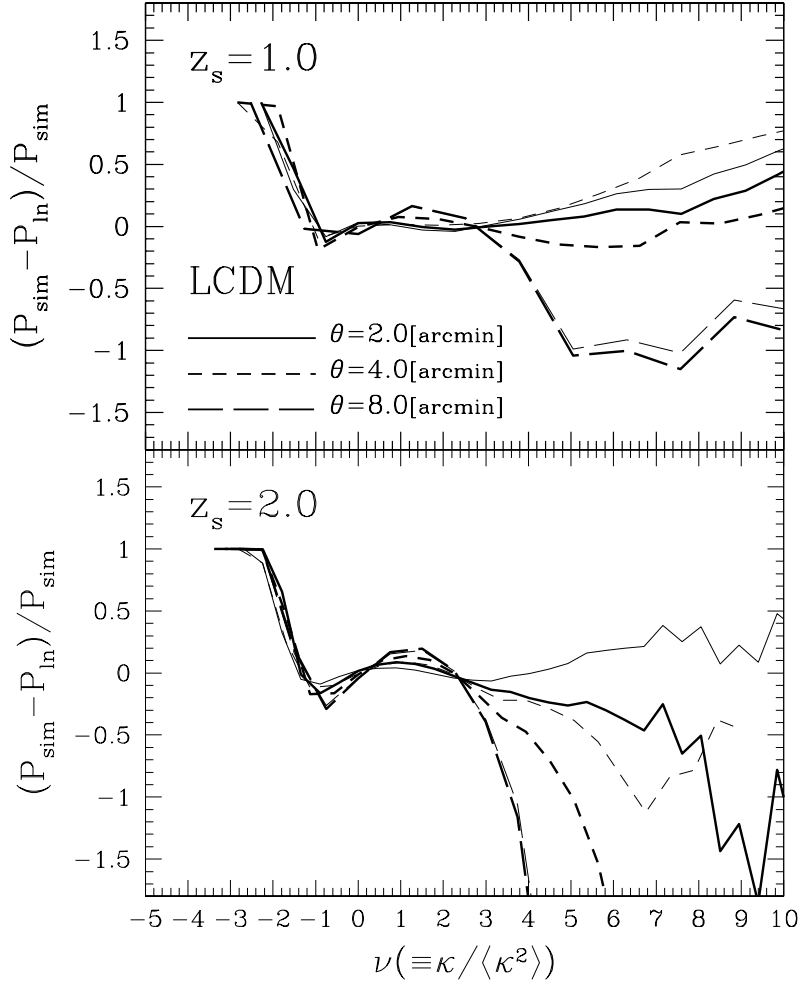}
  \end{center}
    \caption{Differences of one-point PDFs normalized by the simulated PDF, 
      $[P_{\rm sim}(\kappa)-P_{\rm ln}(\kappa)]/P_{\rm sim}(\kappa)$ 
      as a function of level threshold $\nu=\kappa/\var^{1/2}$ 
      in LCDM model. Upper(Lower)-panel shows the results fixing the 
      source redshift to $\zs=1.0$ $(2.0)$. The solid, short-dashed and 
      long-dashed lines represent the cases with smoothing 
      angle $\theta=2.0$', $4.0$' and $8.0$', respectively. In plotting 
      the ratios, the lognormal prediction adopting 
      the averaged minimum value $\kappamin$ (see Fig.\ref{fig: kmin}) is 
      used in thick lines, while the thin lines represent the results 
      adopting the minimum value of each PDF data. 
      \label{fig: err_pdf}}
\end{figure}

In figures \ref{fig: pdf_z1} and \ref{fig: pdf_z2}, solid lines
represent the lognormal predictions (\ref{eq: logpdf}), whose parameters
$\kappamin$ and $\var$ are directly estimated from simulations.  The
agreement between the lognormal model and the simulation results is
generally good.  In particular, at lower source redshift $\zs=1.0$, the
lognormal PDF accurately describes the non-Gaussian tails in the
high-convergence region up to $\kappa\simlt 10\,\,\var^{1/2}$.  In the
case of higher source redshift  $\zs=2.0$, the discrepancy becomes
evident at larger smoothing scale.  The lognormal PDF over-predicts in
the high-convergence region, and under-predicts in the low-convergence
region.  This discrepancy might be ascribed to the projection effect of 
gravitational lensing (see eq.[\ref{eq: def_kappa}]), since the
lognormal PDF is obtained based on the approximation (\ref{eq:
pdf_kappa_del}), which cannot be validated as increasing the source
redshift $\zs$. Nevertheless, even in that case, a better agreement
between lognormal model and simulations was found at small angular
scales. These features are also seen in figure \ref{fig: err_pdf}, 
where the differences between simulated PDF and lognormal PDF 
normalized by the simulated PDF, 
$[P_{\rm sim}(\kappa)-P_{\rm ln}(\kappa)]/P_{\rm sim}(\kappa)$ are
plotted as a function of level threshold $\nu=\kappa/\var^{1/2}$, 
in the case of LCDM model.

Accurate lognormal-fit in the low-$\zs$ case is amazing and is regarded
as a considerable success.  As a closer look at the non-Gaussian tails,
however, lognormal PDF slightly under-predicts the simulations at
low-density region, $\kappa<0$ (see Fig.\ref{fig: pdf_z1} and upper-panel 
of Fig.\ref{fig: err_pdf}). 
Furthermore, at $\theta=2$', the simulation results generally tend to 
deviate from lognormal PDF in the highly 
non-Gaussian tails $\kappa\simgt 10\,\,\var^{1/2}$, although the prediction  
still remains consistent within the 1-$\sigma$ error.  In particular,
in the case of the OCDM model,  the discrepancy becomes apparent even 
at $\kappa\simgt 8\,\,\var^{1/2}$. 
The systematic deviation seen in non-Gaussian tails might be ascribed to 
the presence of virialized objects. Notice that the angular scale 
$\theta=2'$ corresponds to the effective smoothing scale $R\sim 0.5 h^{-1}\,
\mbox{Mpc}$ at mean redshift $z\sim 0.5$. This indicates that high-$\kappa$ 
value is attained by the light-ray propagating through the high-density 
region in the very massive halos with $M \simgt 10^{14}h^{-1}M_{\odot}$. 
Indeed, highly non-Gaussian tails of convergence PDF sensitively 
depends on the detailed structure of non-linear objects, as pointed out by  
Kruse \& Schneider (2000). They construct an analytic model of one-point PDF 
based on the universal profile of dark matter halos and the Press-Schechter 
theory for halo abundance. Since the virialized halos induce 
the highly non-Gaussian tails of the local convergence,  
their treatment would be helpful to describe the non-Gaussian
tails of PDF. In contrast, due to the lack of physical bases, 
no reliable prediction is expected from the empirical lognormal model.

Another reason for discrepancy might be the choice of the
parameter, $\kappamin$.  Strictly speaking, minimum value of $\kappa$
seen in the simulated PDF represents a rare event for the whole data
sets, which are not rigorously equivalent to the averaged value of 40
realization data shown in figure \ref{fig: kmin}. We have also examined
the lognormal-fit adopting the minimum value of the PDF data for
$\kappamin$. The thin-lines in figure \ref{fig: err_pdf} show the results 
in LCDM case adopting the actual minimum value $\kappa$ of each PDF data. 
The results seem somehow improved at small 
angle $\theta=2$'$\sim4$' in high-$\zs$ cases,  
but we rather recognize the fact that lognormal-fit to 
the high-density region is sensitive to the choice of $\kappamin$.

Except for these details, lognormal model of one-point PDF remains a 
fairly accurate model of the convergence field, at least, up to
$\kappa\sim5\var^{1/2}$ and is indeed applicable even at small angular 
scales such as  $2$'$\simlt\theta\simlt4$', irrespective of the 
assumption, (\ref{eq: cumulant_approx}) or (\ref{eq: pdf_kappa_del}).

\subsubsection{Minkowski functionals} 
\label{subsec: minkowski_sim}

Having recognized the successful lognormal-fit to 
the one-point PDF, we next investigate the lognormal model of 
Minkowski functionals. For this purpose, we restrict our analysis to the 
source redshift $\zs=1.0$. Here, the Minkowski functionals for a simulated 
convergence map are calculated by the method developed by 
Winitzki \& Kosowsky (1997).

\begin{figure}
  \begin{center}
    \noindent
    \epsfxsize=15.5cm 
    \epsfbox{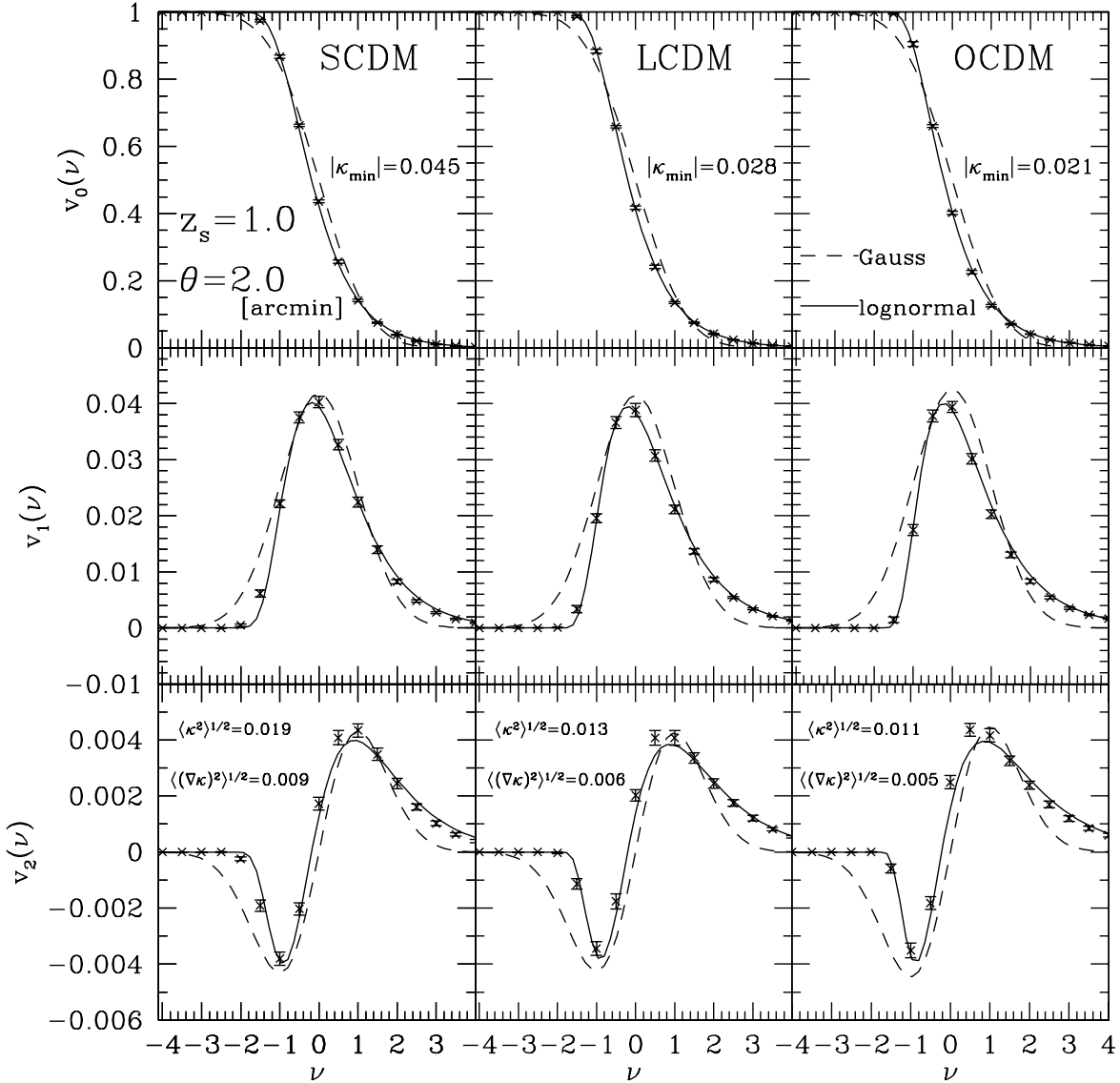}
  \end{center}
  \caption{Minkowski functionals as a function of level threshold 
    $\nu=\kappa/\var$ at the smoothing angle $\theta=2.0$'. 
    The source redshift is fixed to $\zs=1.0$.  Solid lines 
    show the lognormal predictions based on the formulae 
    (\ref{eq: v0_log}),(\ref{eq: v1_log}) and (\ref{eq: v2_log}), where 
    all the parameters are estimated from simulations. The dashed lines 
    are the Gaussian predictions obtained from 
    (\ref{eq: v0_g}), (\ref{eq: v1_g}) and (\ref{eq: v2_g}):   
    SCDM model ({\it left}); LCDM model ({\it middle}); 
    OCDM model ({\it right}). 
    \label{fig: minkow_s2}}
\end{figure}

Figure \ref{fig: minkow_s2} shows the results 
in various CDM models fixing the 
smoothing angle $\theta=2.0$'. 
The Minkowski functionals are plotted 
against the level threshold $\nu=\kappa/\var^{1/2}$ with the interval, 
$\Delta\nu=0.5$, so that data in each bin is approximately regarded
as statistically independent. 
The solid lines depict the lognormal predictions 
(\ref{eq: v0_log}), (\ref{eq: v1_log}) and (\ref{eq: v2_log}).

Similar to the one-point PDF,  
a significant non-Gaussian signature is detected from 
the asymmetric shape of the Minkowski functionals, especially 
from the Euler characteristic $v_2$. In marked contrast with the 
Gaussian predictions, the lognormal predictions 
remarkably reproduce the simulation results, not only the 
shape dependences but also the amplitudes. 
The agreement between lognormal prediction and simulations 
still remains accurate over the rather wider range, $-4<\nu <4$,  
where the discrepancy seen in the one-point 
PDF of the OCDM model is not observed. 
Since the prediction has no adjustable parameter and 
only uses the information from an output data, 
this agreement is successful.

\begin{figure}
  \begin{center}
    \noindent
    \epsfxsize=15.5cm 
    \epsfbox{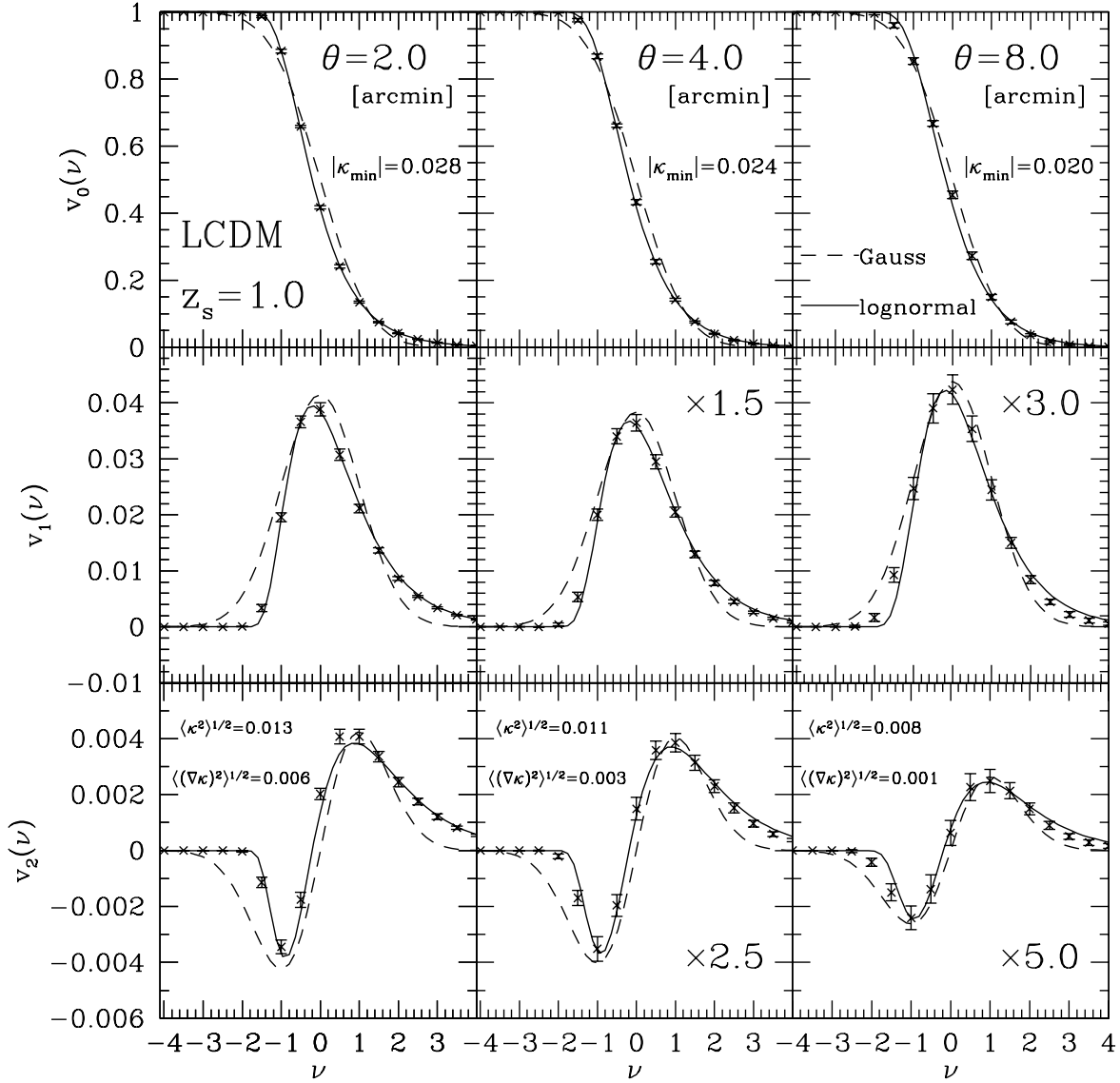}
  \end{center}
  \caption{Same as Fig.\ref{fig: minkow_s2}, but we fix
    the cosmological model to LCDM model. 
    Here, for illustrative purpose, the amplitude of $v_{1}$ and $v_{2}$ 
    at the smoothing angle $\theta=4,\,8$ [arcmin] are enhanced in 
    order to clarify the differences as indicated in each panel: 
    $\theta=2$' ({\it left}); $\theta=4$' ({\it middle}); 
    $\theta=8$' ({\it right}). 
\label{fig: minkow_lcdm}}
\end{figure}

Figure \ref{fig: minkow_lcdm} depicts the results with various smoothing
angles fixing the cosmology to LCDM model.  For illustrative purpose, 
the amplitudes of $v_1$ and $v_2$ at the 
smoothing angle $\theta=4$' and $8$' are artificially changed by
multiplying the factors as indicated in each panel, 
in order to make the non-Gaussianity manifest.  
The Minkowski functionals tend to approach the Gaussian prediction as
increasing the smoothing angle, consistent with the behaviors of
one-point PDF.  The results in other cosmological models are also
similar and the agreement between lognormal models and simulations is
satisfactory.

To manifest the accuracy of the lognormal-fit in contrast to 
the other existing analytical models, let us now consider the 
perturbation predictions. 
Employing the Edgeworth expansion, 
  the perturbative expressions for Minkowski functional are derived 
  analytically (Matsubara 2000; Sato et al. 2001):
\begin{eqnarray}
v_0(\nu)&\simeq& v_{0,{\scriptscriptstyle\rm G}}(\nu)
+\var^{1/2}\, \frac{1}{6\sqrt{2\pi}}\,e^{-\nu^2/2}\,S_3^{(0)}\,H_2(\nu), 
\label{eq: v0_Edge}
\\
v_1(\nu)&\simeq&v_{1,{\scriptscriptstyle\rm G}}(\nu) 
\left[ 1+ \var^{1/2}\left\{\frac{S_3^{(0)}}{6}H_3(\nu)
+\frac{S_3^{(1)}}{3}H_1(\nu)\right\}\right],
\label{eq: v1_Edge}
\\
v_2(\nu)&\simeq& v_{2,{\scriptscriptstyle\rm G}}(\nu) 
\left[1+ \frac{\var^{1/2}}{H_{0}(\nu)}
\left\{\frac{S_3^{(0)}}{6}H_4(\nu)+\frac{2S_3^{(1)}}{3}H_2(\nu)+
  \frac{S_3^{(2)}}{3}\right\}\right]. 
\label{eq: v2_Edge}
\end{eqnarray}
Notice that in the case of the three-dimensional density field, the above 
expansion is valid up to the RMS of $\delta$, $\sigma\simlt0.3$ 
(Matsubara \& Yokoyama 1996; Matsubara \& Suto 1996),  
which can be translated into the condition, 
$\var^{1/2}\simlt0.3\,\,|\kappamin|$.  
Here, the function $H_{n}(\nu)$ denotes $n$-th order Hermite polynomial, 
$H_{n}(x)\equiv(-1)^{n}e^{x^{2}/2}(d/dx)^{n}e^{-x^{2}/2}$,  and   
the quantities $S_3^{(i)}$ represent the skewness 
parameters defined as 
\begin{eqnarray}
  \label{eq: S_0}
  S_3^{(0)} &=& \frac{\langle\kappa^3\rangle}{\var^2},
\\
  \label{eq: S_1}
  S_3^{(1)} &=& -\frac{3}{4}\,
  \frac{\langle\kappa^{2}\cdot\nabla^{2}\kappa\rangle}{\var\,\derivar},
\\
  \label{eq: S_2}
  S_3^{(2)} &=& -3\,
  \frac{\langle(\nabla\kappa\cdot\nabla\kappa)\nabla^{2}\kappa\rangle}
  {\derivar^{2}}.
\end{eqnarray}
Equations (\ref{eq: v0_Edge})-(\ref{eq: v2_Edge}) imply that, 
in the weakly nonlinear regime, the non-Gaussian features on the
Minkowski functionals can be completely described by the above
parameters, which can be evaluated by the second-order perturbation
theory of structure formation (e.g., Bernardeau et al. 1997).  
It is worth noting that in the case of the lognormal model, all the
skewness parameters, $S_3^{(\alpha)}$ are equal to $3$ 
(Hikage, Taruya \& Suto 2001).

\begin{figure}
  \begin{center}
    \noindent
    \epsfxsize=15.5cm 
    \epsfbox{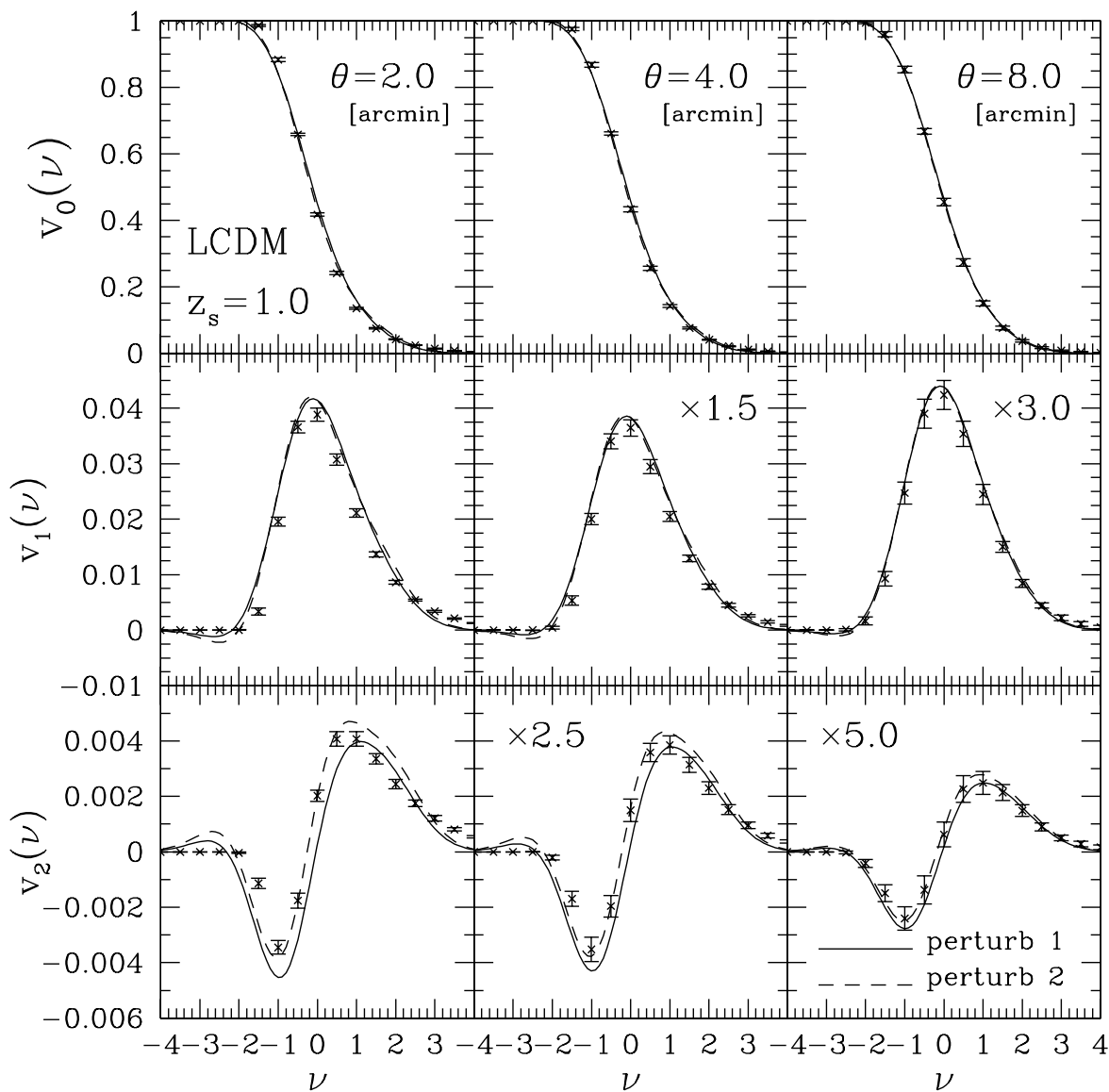}
  \end{center}
  \caption{Edgeworth expansion of Minkowski functionals compared with 
    the simulations. In each panel, the results of $\zs=1$ case in 
    LCDM model are shown. Solid lines ({\it perturb 1}) indicate the 
    perturbation results, where the 
    skewness parameters are calculated via perturbation theory, while 
    the skewness parameters of the dashed lines ({\it perturb 2}) 
    are estimated from simulation data directly. In plotting both cases,  
    the variances $\var$ and $\derivar$ are fitted to the numerical 
    simulations. Note that the amplitude of $v_{1}$ and $v_{2}$ at the 
    smoothing angle $\theta=4,\,8$ [arcmin] are enhanced in order to 
    clarify the differences as indicated in each panel. 
    \label{fig: minkow_perturb}
    }
\end{figure}

\begin{figure}
  \begin{center}
    \noindent
    \epsfxsize=15.5cm 
    \epsfbox{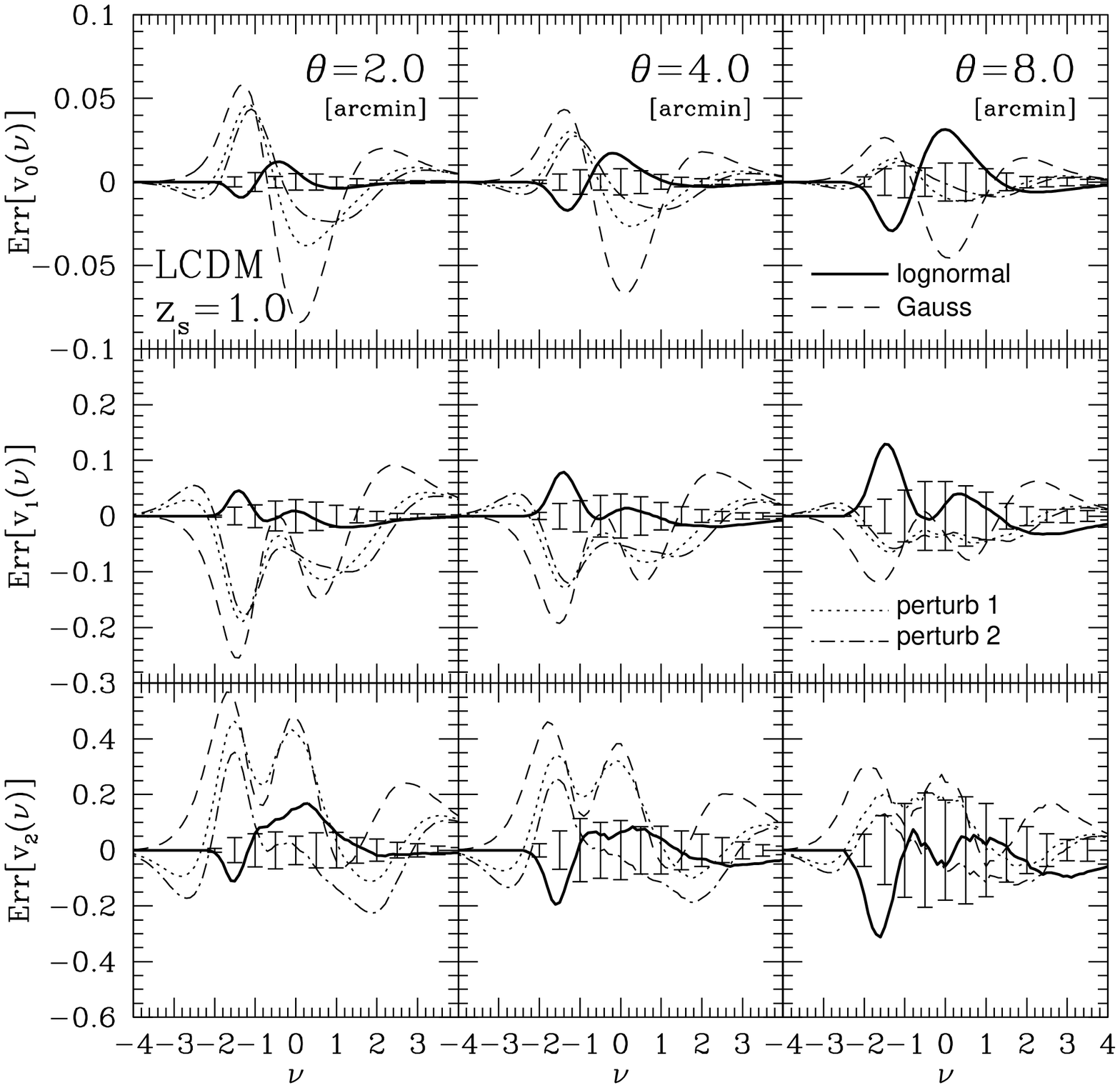}
  \end{center}
  \caption{Fractional errors of various model predictions in 
    Minkowski functionals,  
    $\mbox{Err}[v_i(\nu)]\equiv[v_{i}^{\rm\scriptscriptstyle(sim)}(\nu)-
    v_{i}^{\rm\scriptscriptstyle(model)}(\nu)]/
    v_{i,{\rm\scriptscriptstyle max}}^{\rm\scriptscriptstyle (sim)}$, 
    in the case of LCDM model with $\zs=1$. Here, 
    $v_{i,{\rm\scriptscriptstyle max}}^{\rm\scriptscriptstyle (sim)}$
    denotes the maximum value of $v_i(\nu)$ estimated from 
    simulation. In each panel, the fractional errors for the 
    lognormal and Gaussian prediction are plotted as thick-solid and 
    dashed lines, respectively. The dotted- and dot-dashed lines 
    represent the perturbative predictions based on the Edgeworth 
    expansion (dotted: {\it perturb 1}; dot-dashed: {\it perturb 2}). 
    In each panel, the error-bars around zero mean indicate the 
    realization error of ray-tracing simulations: $\theta=2'$ ({\it left}); 
    $\theta=4'$ ({\it middle}); $\theta=8'$ ({\it right}). 
    \label{fig: error_minkow}
    }
\end{figure}

Figure \ref{fig: minkow_perturb} plots the perturbation 
results in LCDM model. The source redshift is fixed to $\zs=1.0$. 
The solid lines represent the results in which the skewness parameters are 
evaluated using the second-order perturbation theory ({\it perturb 1}), 
while the dashed lines depict the results using those estimated from 
simulations ({\it perturb 2}). In both cases, the variances $\var$ and 
$\derivar$ in the expressions (\ref{eq: v0_Edge})-(\ref{eq: v2_Edge}) 
are estimated from simulations. 
Also, in figure \ref{fig: error_minkow}, 
the comparison between various model predictions is summarized, 
introducing the fractional error, defined by 
\begin{equation}
  \label{eq: error_mink}
 \mbox{Err}[v_i(\nu)]\equiv   \frac{v_i^{\rm\scriptscriptstyle(sim)}(\nu)-
v_i^{\rm\scriptscriptstyle(model)}(\nu)}
{v_{i,{\rm max}}^{\rm\scriptscriptstyle(sim)}}, 
\end{equation}
with the quantity $v_{i,{\rm max}}^{\rm\scriptscriptstyle(sim)}$ being
the maximum value of $v_i(\nu)$ among the mean values of the simulation 
for each Minkowski functional: 
$v_{0,{\rm max}}^{\rm\scriptscriptstyle(sim)}=1.0$,  
$v_{1,{\rm max}}^{\rm\scriptscriptstyle(sim)}=0.040$ and 
$v_{2,{\rm max}}^{\rm\scriptscriptstyle(sim)}=0.0042$ at 
angular scale $\theta=2'$, for instance.  
As a reference,
the 1-$\sigma$ error of simulation results is plotted as error-bars
around zero mean in each panel.

At the large smoothing angle $\theta=8$', both of the perturbation
results tend to reconcile with each other and fit to the 
simulation results well within the 1-$\sigma$ error. As the smoothing angle 
decreases, however, the perturbation results cease to fit the simulations, 
because the RMS of local convergence reach 
$\var^{1/2}\simgt0.4\,\,|\kappamin|$ and the Edgeworth expansion breaks 
down. Furthermore, the second-order perturbations under-predict the skewness 
parameters, compared with those estimated from simulations, which lead to 
the different predictions (compare {\it perturb 1}  with {\it perturb 2} in 
left- and middle-panels in 
Figs.\ref{fig: minkow_perturb}, \ref{fig: error_minkow}). 
In particular, the circumference $v_{1}$ shows a 
peculiar behavior, $v_1<0$, which is not allowed by definition. 
Notice that even in these cases, the fractional error 
$\mbox{Err}[v_i(\nu)]$ in the lognormal prediction still remains smaller,  
although the systematic deviations in every model are not so large.

From these discussions, we conclude that the empirical lognormal models 
can provide a good approximation for the Minkowski functionals, compared
to the current existing models.

\section{Consistency with higher-order moments}
\label{sec: moments}

  Since the skewness of the local convergence has been proposed as a 
  simple statistical estimator of non-Gaussian signature to 
  determine the cosmological parameters, various authors have
  investigated the usefulness of this quantity using ray-tracing
  simulations.  These analyses have revealed that the skewness at
  small-angle $\theta\simlt10$' exhibits the significant influence of
  nonlinear clustering and a more reliable theoretical model beyond the
  perturbation theory is needed.  According to these facts,
  non-perturbative predictions based on the ``hyper-extended
  perturbation theory''(Scoccimarro \& Frieman 1999) or non-linear
  fitting formula of bi-spectrum(Scoccimarro \& Couchman 2001) are
  examined (Hui, 1998; Van Waerbeke et al. 2001b).

  In general, the one-point PDF as well as the Minkowski functionals 
  characterizes a family of higher-order statistics. 
  Hence, as a consistency check of the lognormal prediction, 
  it seems natural to analyze the higher-order moments of local 
  convergence. From an empirical lognormal PDF (\ref{eq: logpdf}), 
  the skewness and the kurtosis of the local convergence 
  defined by $S_{3,\kappa}\equiv\langle\kappa^{3}\rangle/\var^{2}$ 
  and $S_{4,\kappa}\equiv(\langle\kappa^{4}\rangle-3\var^{2})/\var^{3}$ 
  are respectively given by 
\begin{equation}
  \label{eq: skewness_ln}
  S_{3,\kappa} = \frac{1}{|\kappamin|}
  \left(3+\frac{\var}{|\kappamin|^2}\right), 
\end{equation}
\begin{equation}
  \label{eq: kurtosis_ln}
  S_{4,\kappa} = \frac{1}{|\kappamin|^2}
  \left\{16+15\frac{\var}{|\kappamin|^2}+
6\left(\frac{\var}{|\kappamin|^2}\right)^{2}+
\left(\frac{\var}{|\kappamin|^2}\right)^{3}\right\}.   
\end{equation}

\begin{figure}
  \begin{center}
    \noindent
    \epsfxsize=11.0cm 
    \epsfbox{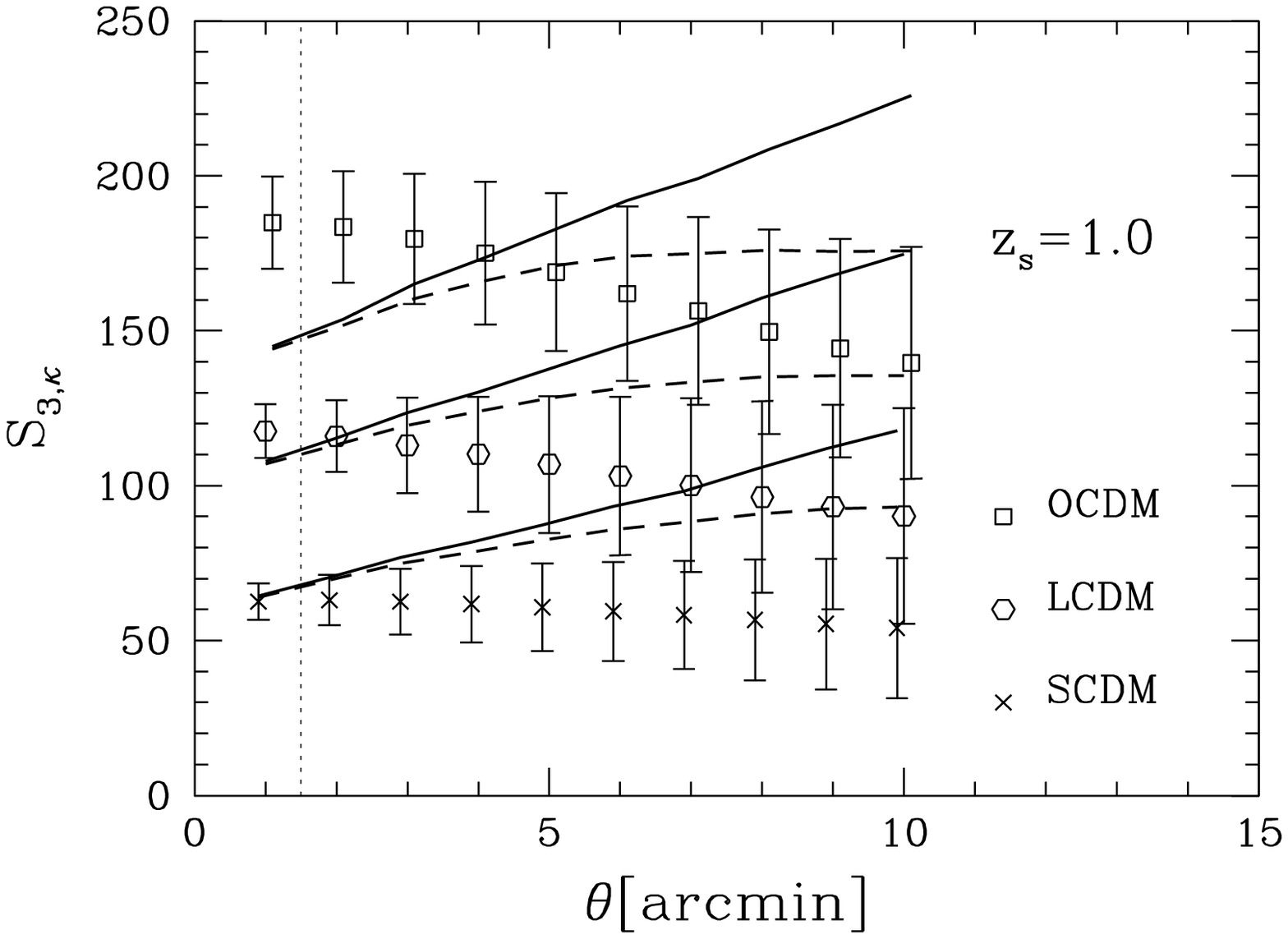}

\vspace*{1.5cm}

    \noindent
    \epsfxsize=11.0cm 
    \epsfbox{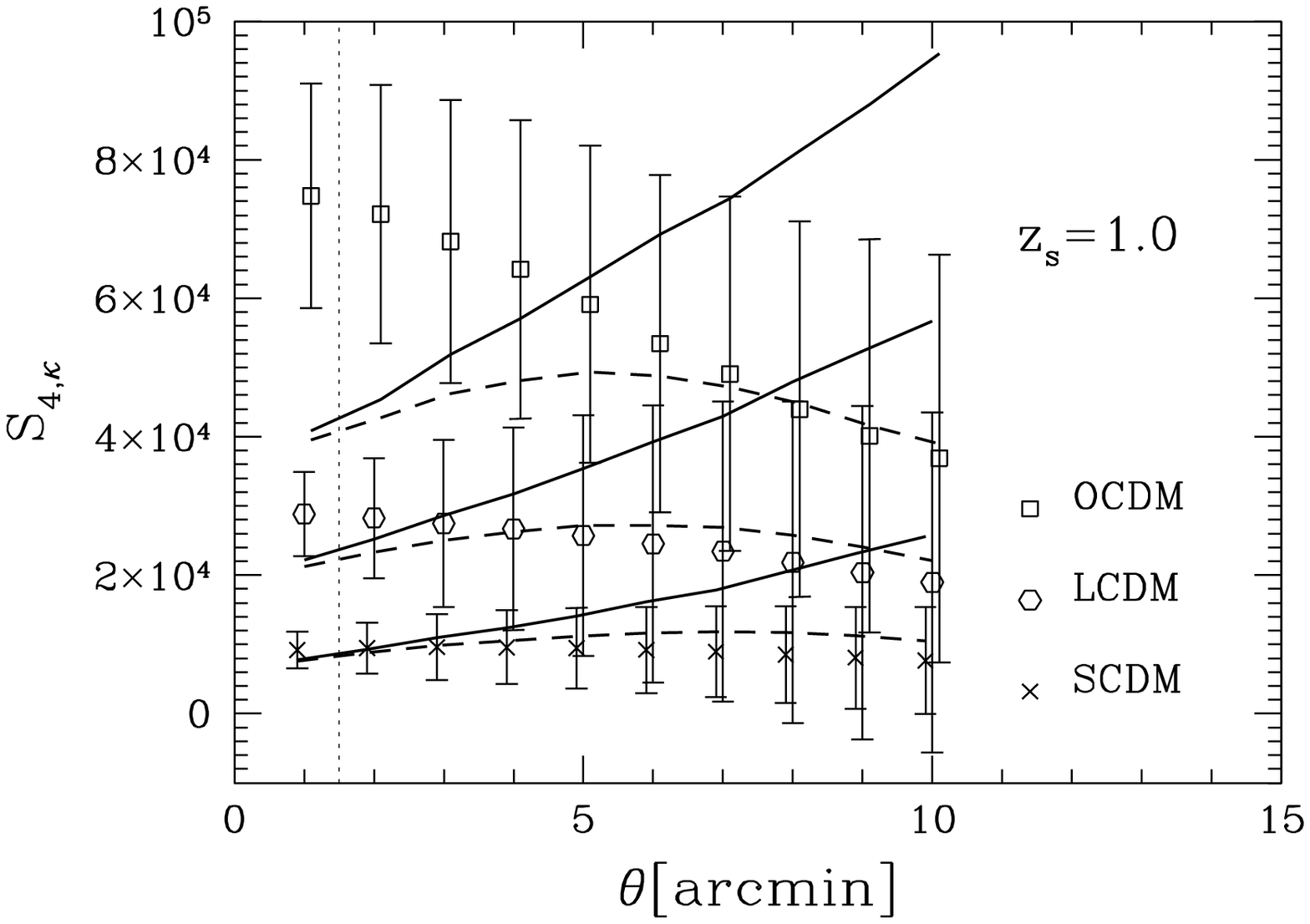}
  \end{center}
  \caption{Skewness $S_{3,\kappa}$({\it upper}) and kurtosis 
    $S_{4,\kappa}$({\it lower}) 
    as a function of smoothing angle. The crosses, hexagons and  
    squares with error-bars represent the simulation results in 
    SCDM, LCDM and OCDM, respectively. The solid lines show the 
    lognormal prediction based on the expressions, 
    (\ref{eq: skewness_ln}) and (\ref{eq: kurtosis_ln}). 
    The dashed lines also indicate the lognormal prediction, 
    but here we take into account the limited range of 
    convergence data, i.e., $\kappamin<\kappa<\kappa_{\rm max}$. 
    \label{fig: skewness}}
\end{figure}

Figure \ref{fig: skewness} shows the direct measurement of $S_{3,\kappa}$ 
and $S_{4,\kappa}$ fixing the source redshift to $\zs=1.0$. 
The error-bars indicates 
1-$\sigma$ error estimated from the 40 realization data.  
The lognormal predictions (\ref{eq: skewness_ln}) and (\ref{eq: kurtosis_ln})
are depicted as solid lines. 

In practice, the convergence field $\kappa$ in numerical simulation 
does not extend the entire range between $\kappamin$ and $+\infty$, 
but is limited as $\kappamin<\kappa<\kappamax$, due to the 
finite sampling effect from a limited size of simulation data (see 
Kayo et al. 2001 in the case of three-dimensional density field $\delta$). 
Thus, the $n$-th order moments of $\kappa$ given by 
\begin{equation}
  \label{eq: n-th}
  \langle\kappa^{n}\rangle=\int_{\kappamin}^{\kappamax}
\,\,d\kappa\,\,P_{\rm ln}(\kappa)\,\,\kappa^{n}
\end{equation}
may provide a better description in evaluating the skewness $S_{3,\kappa}$ and 
the kurtosis $S_{4,\kappa}$.

The dashed lines in figure \ref{fig: skewness} show the lognormal 
prediction based on the equation (\ref{eq: n-th}), where 
the cutoff parameter $\kappamax$ is adopted as the averaged maximum 
value from each realization data. 
In contrast to the accurate fit seen in the 
one-point PDF, the lognormal-fit of skewness and kurtosis seems very poor. 
The prediction without cutoff $\kappamax$ tends to over-predict as increasing 
smoothing angle, which shows opposite behavior compared to the simulations.  
As for the lognormal model based on (\ref{eq: n-th}), 
while the predictions reduce their amplitudes and broadly agree with 
simulations in the case of $S_{4,\kappa}$, 
they still exhibit some systematic discrepancies in $S_{3,\kappa}$. We 
have also examined the lognormal-fit adopting the 
minimum and maximum values of PDF data itself, but 
the result is not improved drastically.

At first glance, these discrepancies seem to contradict with the results
in one-point PDF, however, a closer look at one-point PDF reveals that
the lognormal PDFs slightly under-predict the simulation results at the
low-density region. This tiny discrepancy may 
be ascribed to the overestimation of skewness. In other words, the 
skewness as well as kurtosis is very sensitive to the rare events, i.e, 
high- and low-convergence parts of the non-Gaussian tails. This sensitivity 
is clearly shown in the OCDM model. The simulated PDF at smoothing angle
$\theta=2$' exhibits a highly non-Gaussian tails and it overshoots the
lognormal prediction (see upper-right panel in Fig.\ref{fig: pdf_z1}). 
The resultant skewness and kurtosis 
yield values larger than those of lognormal prediction.  
On the other hand, in SCDM and LCDM models, lognormal PDFs accurately 
fit to the highly non-Gaussian tails and thereby the predictions of 
skewness and kurtosis becomes relatively consistent with simulations, 
at least on the small scales, $2$'$\leq\theta\leq4$'.

Therefore, the lognormal model of convergence does not provide an 
accurate prediction for the statistics sensitive to the rare events. 
This disagreement simply reflects the fact that the empirical 
lognormal model does not correctly describe projected structure of 
dark matter halos. On the other hand, a sophisticated 
non-perturbative model based on the hyper-extended perturbation 
theory or fitting formula of bi-spectrum is constructed so as to 
reproduce the N-body results of higher-order moments, which can 
provide an accurate prediction for the convergence skewness 
(Hui, 1998; Van Waerbeke et al. 2001b). Hence, for the prediction of 
higher-order moments,  the non-perturbative model is more useful 
and reliable than the lognormal model. In contrast, the lognormal model 
fairly describes non-Gaussianity around the peak of the PDFs over 
the broad range, $\kappa\simlt 5\,\var^{1/2}$,   
which cannot be described by such a non-perturbative model. Thus, at least 
as a complementary approach, the lognormal model is useful 
beyond the perturbative theory, and applicable in characterizing the 
one-point PDF and the Minkowski functionals.

\section{Conclusion and Discussion}
\label{sec: summary}
%
%
%
%
%
In the present paper, we have quantitatively investigated the extent 
to which the lognormal model fairly describes 
the statistics of weak lensing fields on linear and nonlinear scales 
using the ray-tracing simulations. 
The validity of lognormal model has been checked in details 
by comparing the lognormal predictions of 
 the one-point PDF and the Minkowski functionals of convergence field
with their simulation results. 

The convergence field seen in the one-point PDF and the Minkowski
functionals displays the non-Gaussian feature and significantly deviates
from the Gaussian predictions. We have shown that the analytic formulae
for lognormal models are useful and accurately describe the simulation
results on both small and large smoothing angular scales in the case of
the low-$\zs$ data, while the perturbative prediction by Edgeworth
expansion fails to reproduce the simulation results on small scales
because of the nonlinearity of the underlying three-dimensional density
field. The detailed comparison revealed that the lognormal model does
not provide an accurate prediction for the statistics sensitive to the
rare events such as the skewness and kurtosis of the convergence.  We
therefore conclude that, as long as we are concerned with the
appropriate range of the convergence, $\kappa\simlt
5\,\var^{1/2}$, the lognormal model empirically but quantitatively gives
a useful approximation characterizing the non-Gaussianity features on 
the convergence field.

The results obtained here will lead to an
important improvement for the estimation of cosmological parameters
using the Minkowski functionals (Sato, Takada, Jing \& Futamase
2001). Although the original methodology has been proposed with the use
of the Edgeworth formulae, (\ref{eq: v0_Edge})-(\ref{eq: v2_Edge}), a
more reliable estimation of cosmological parameters will be possible
using the lognormal models.  
In the light of this, a reliable theoretical prediction for 
the model parameters of the lognormal formulae 
$\kappamin$, $\var$ and $\derivar$ should be further explored. 
As shown in figures \ref{fig: kmin} and
\ref{fig: varkappa}, the simulation results reveal that $\kappamin$ is
very sensitive to the density parameter $\Omega_0$, while the $\var$ and
$\derivar$ depend on $\Omega_0$ and $\sigma_8$.  
Such predictions will offer a new opportunity to determine the cosmological
parameters, independently of the method using the higher-order moment,
$S_{3,\kappa}$. In particular, if we do not focus on the overall
normalization of the Minkowski functionals $v_1$ and $v_2$ controlled by
the $\var$ and $\derivar$, the non-Gaussian features on those
functionals are primarily controlled by $\kappamin$, and, for example,
the intuitive way to predict $\kappamin$ discussed in Appendix A could
be used in constraining $\Omega_{0}$. Of course, for proper comparison
with observation, there are other systematic effects we have to take
into account. One is the redshift distribution of source galaxies. 
Another important effect is the contamination by the intrinsic source 
ellipticity. This not only reduces the signal-to-noise ratio, but also 
systematically affects the non-Gaussianity of the weak lensing signal 
(Jain, Seljak \& White 2000). These effects should be correctly 
incorporated into the lognormal predictions.  These issues are now in 
progress, and will be presented elsewhere.

The statistics of cosmic shear directly reflect the statistical feature
of mass distribution and using this fact, one might even discriminate
the nature of dark matter. Furthermore, the weak lensing statistics have
a potential to reveal the nonlinear and stochastic properties of galaxy
biasing.  In any cases, we expect that the lognormal property of the
weak lensing field is helpful and plays an important role in extracting
the various cosmological information.

\bigskip
\bigskip

We thank J.Sato for providing us the code to calculate the Minkowski 
functionals from the ray tracing simulation data. 
We also thank Y.Suto for careful reading of the manuscript and 
critical comments, C.Hikage for invaluable discussion, 
T.Buchert and M.Bartelmann for useful comments. I.K is supported by 
Takenaka-Ikueikai Fellowship. T.H. and M.T acknowledge supports from 
Japan Society for Promotion of Science (JSPS) Research Fellowships. 

\clearpage
\appendix 
\section{Influence of finite sampling on estimation of $\kappamin$}
\label{app: kappamin}
%
%
%
%
Minimum value of the convergence, $\kappamin$ is theoretically 
expected to be equal to $\kappaempty$ (eq.[\ref{eq: empty}]). 
In practice, however,   
the minimum value estimated from the simulation data 
can be systematically larger than those obtained from the empty beam, 
due to the finite sampling from the limiting survey area.

To see the influence of the finite sampling effect, 
let us roughly estimate the minimum value, $\kappamin$. Assuming that 
the {\it prior} one-point PDF, $P^{\rm\scriptscriptstyle (prior)}(\kappa)$ 
is approximately described by the lognormal PDF, 
in which the minimum value is characterized by $\kappaempty$ instead of 
the actual value $\kappamin$: 
\begin{equation}
  \label{eq: prior_pdf}
P^{\rm(prior)}(\kappa)\,\,=
\frac{1}{\sqrt{2\pi}\sigmaln}\,\,\exp
\left\{-\frac{[\ln(1+\kappa/|\kappaempty|)+\sigmaln^2/2]^2}{2\sigmaln^2}
\right\}\,\frac{d\kappa}{\kappa+|\kappaempty|}, 
\end{equation}
with the quantity $\sigmaln^{2}$ being 
$\ln\left(1+\var/|\kappaempty|^{2}\right)$.
Then the probability that the observed minimum value $\kappamin$ 
systematically deviates from $\kappaempty$ is given by 
\begin{equation}
 \int_{-|\kappaempty|}^{-|\kappamin|} 
d\kappa\,\,P^{\rm\scriptscriptstyle (prior)}(\kappa) = 
\frac{\pi\thetath^2}{\theta_{\rm field}^2}, 
\label{eq: kmin}
\end{equation}
where $\thetath$ is the top-hat smoothing angle, 
$\theta_{\rm field}$ is field-of-view angle. The right hand side of 
equation (\ref{eq: kmin}) represents the lower limit of probability 
determined from the expectation number of independent sampling area. 
The minimum value $\kappamin$ is obtained by solving equation 
(\ref{eq: kmin}). 

Based on the equation (\ref{eq: kmin}), 
the resultant values of $\kappamin$ are summarized 
in figure \ref{fig: kmin} ({\it thin-lines}).  
Here, the field-of-view angle $\theta_{\rm field}$ is fixed to $5^{\circ}$   
and the parameters in the prior lognormal PDF, $\var$ and $\kappaempty$ 
are computed according to the theoretical predictions  
(\ref{eq: var_kappa}) and (\ref{eq: empty}), respectively.  
%
%
%
%
%
%
%
\section{Effect of finite resolution and variances $\var$ and $\derivar$}
\label{app: var_derivar}
%
%
%
%
Statistical analysis based on the N-body simulation with 
PM algorithm should be carefully treated if we deal with 
the statistics on small scales.    
In our ray-tracing simulation, 
the box size of each simulation, $L_{\rm box}(z)$,  
are determined so as to satisfy the resolution angle  
$\theta_{\rm res}\approx 1.5$', i.e, 
$L_{\rm box}(z)\approx r(\chi(z))\,\theta_{\rm res}$. 
The mesh of the PM algorithm is fixed to 
$252^{2}\times 512$ in each simulation box.  Among these 
parameters, the number of mesh severely restricts the Fourier modes 
of mass fluctuations, which could affect the evaluation of $\var$ and 
$\derivar$, depending on the choice of smoothing filter.

In order to investigate the significance of finite mesh-size,  
the theoretical predictions (\ref{eq: var_kappa}) and 
(\ref{eq: var_nablakappa}) are modified according to 
the PM N-body code. Since the influence of finite mesh-size 
mainly affects the high-frequency part of the fluctuations, 
the Fourier-integrals in the expressions (\ref{eq: var_kappa}) 
and (\ref{eq: var_nablakappa}) are modified to 
\begin{equation}
  \label{eq: discretize}
  \int_{0}^{\infty} d\kperp\,\,\,\Delta^{2}(\kperp)
W_{\rm TH}^{2}(\kperp r \theta) 
\,\,\,\,\, \Longrightarrow\,\,\,\,\,
\sum_{n=1}^{N_{\rm max}} \Delta \kperp\,\,\,\Delta^{2}
(k_{\perp,n})\,W_{\rm TH}^{2}(k_{\perp,n}\,r\theta),
\end{equation}
where 
\begin{equation}
  \label{eq: Delta}
  \Delta^{2}(\kperp) \equiv\left\{
\begin{array}{cl}
(\kperp/2\pi)\,\,P_{\rm\scriptscriptstyle mass}(\kperp)
& \,\,\,\,\,\,\,\,\,\,\,\,\mbox{for} \,\,\,\var,
\\
\\
(\kperp^3/2\pi)\,\,P_{\rm\scriptscriptstyle mass}(\kperp)
& \,\,\,\,\,\,\,\,\,\,\,\,\mbox{for} \,\,\,\derivar.
\end{array}
\right. 
\end{equation}
The $n$-th Fourier mode $k_{\perp,n}$ is  given by 
$k_{\perp,n}=\Delta\kperp\times n$ and 
the interval $\Delta\kperp$ is set to 
$\Delta\kperp=2\pi/L_{\rm box}(z)$.

In the above modification, the number of 
Fourier modes, $N_{\rm max}$ might be crucial in evaluating the 
quantity sensitive to the high frequency mode, 
which is related to the number of mesh, 
$N_{\rm mesh}=256$. Recall that the Nyquist frequency 
restricts the high-frequency mode to 
$k_{\rm Nyq}=(\Delta\kperp/2)N_{\rm mesh}$, 
which implies $N_{\rm max}=N_{\rm mesh}/2$. Further,  
the number of independent Fourier mode is reduced by the factor 
$1/2$  in evaluating the power spectrum. Thus, the high-frequency cutoff 
is roughly given by 
$k_{\rm cut}\approx \sqrt{2}\times \Delta\kperp(N_{\rm mesh}/4)$, 
which yields 
\begin{equation}
  N_{\rm max}\,\,=\,\,\frac{N_{\rm mesh}}{2\sqrt{2}}\approx 90.  
\end{equation}

\begin{figure}
  \begin{center}
   \leavevmode
\epsfxsize=11.0cm 
\epsfbox{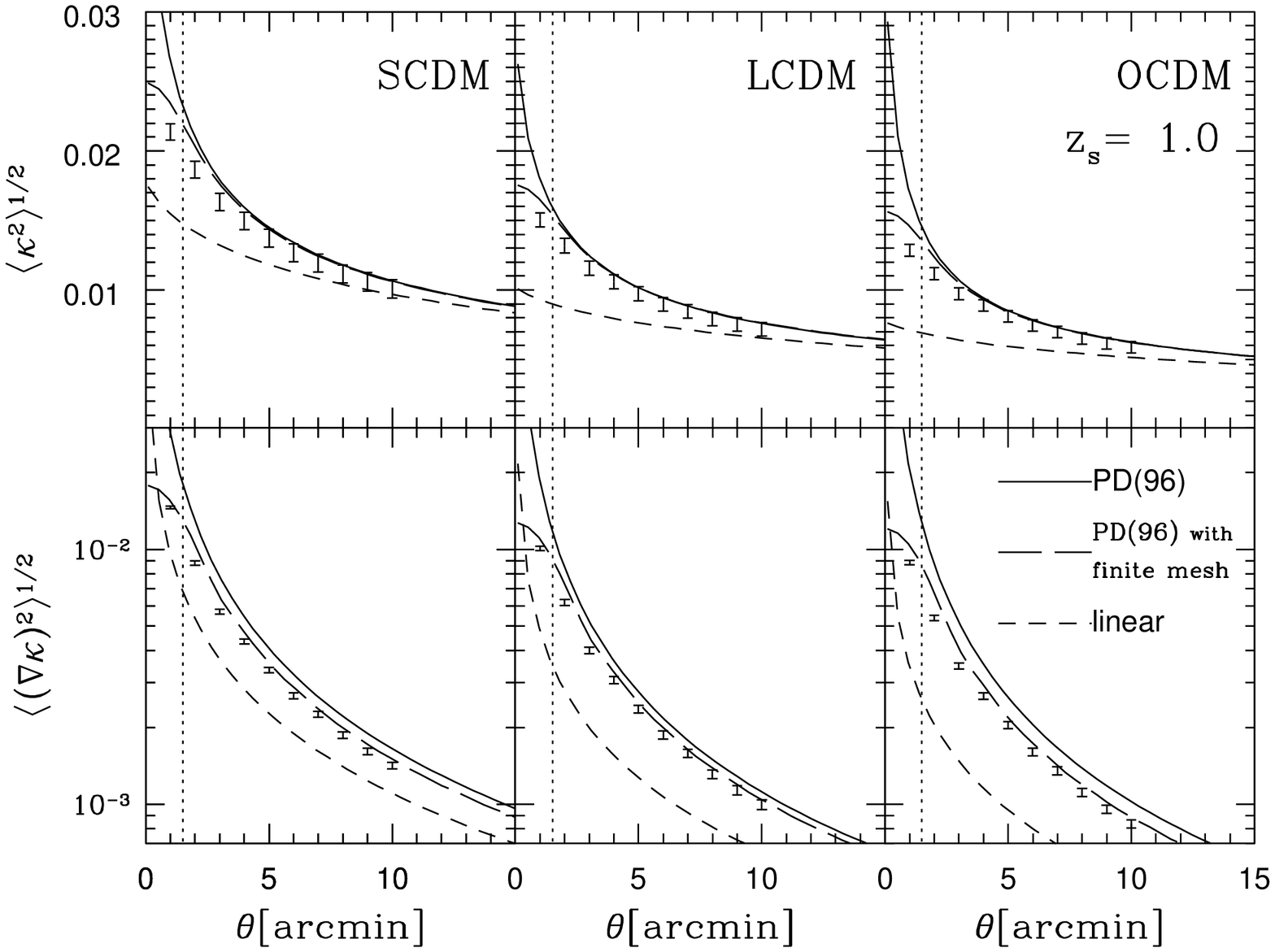}

\vspace*{0.5cm}

\noindent
\epsfxsize=11.0cm 
\epsfbox{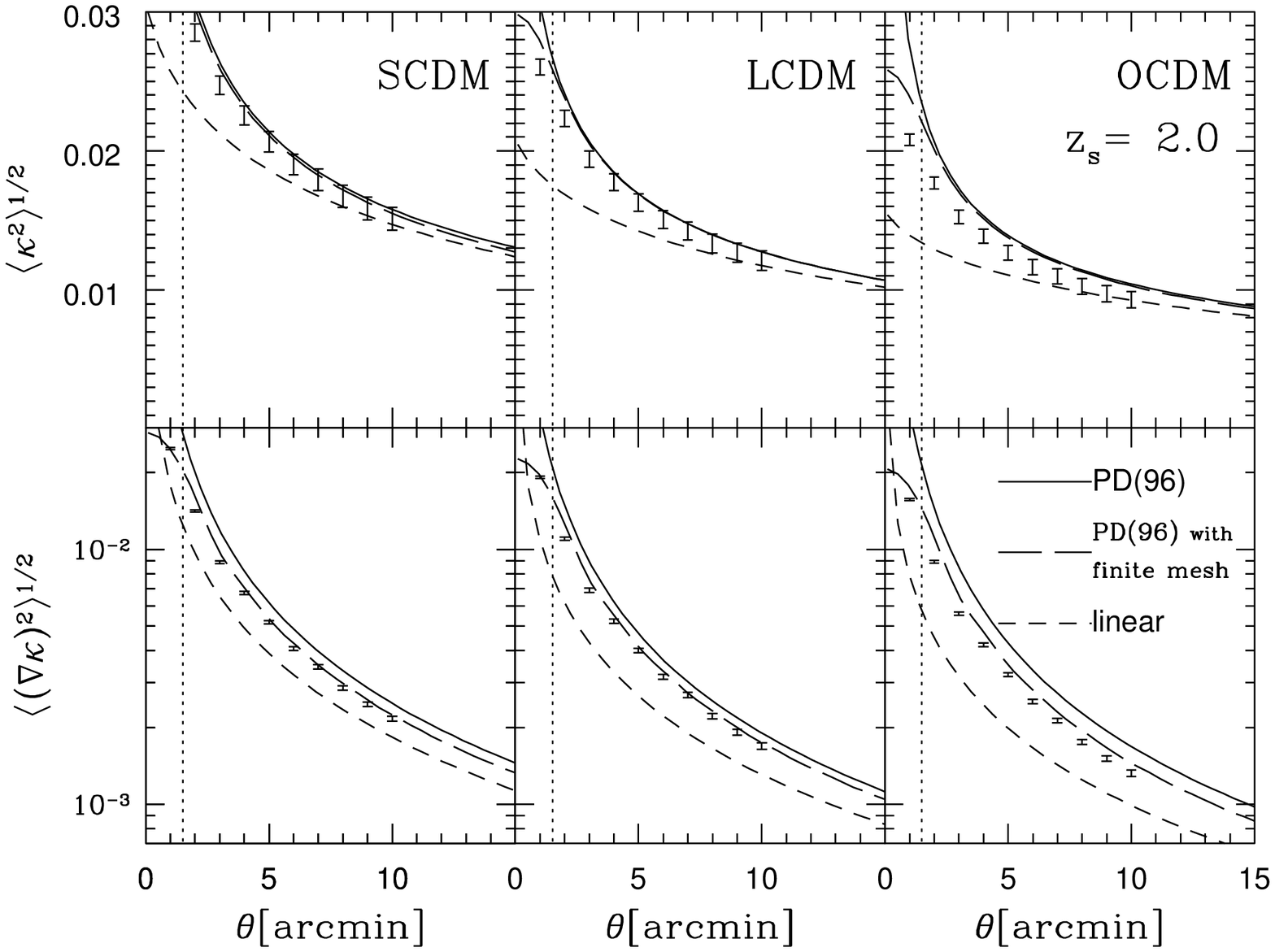}
  \end{center}
\caption{RMS of convergence field({\it upper-panel}) and its gradient field 
  ({\it lower-panel}) as a function of 
  smoothing angle, plotted in linear- and log-scales. 
  The error-bars indicate the simulation results, where error-bars are 
  estimated from the $40$ realization of ray-tracing data 
  by randomly shifting each simulation boxes. The solid and short-dashed lines 
  represent the theoretical prediction based on the 
  expressions (\ref{eq: var_kappa}) and (\ref{eq: var_nablakappa}). 
  In evaluating these equations, nonlinear mass power 
  spectra by Peacock \& Dodds (1996) are adopted in solid lines, while 
  the linear power spectra are used in short-dashed lines. 
  The long-dashed lines also represent the nonlinear prediction, 
  but taking account of the finite resolution of PM N-body simulations 
  (see Appendix B for details): SCDM model ({\it Left}) ; 
  LCDM model ({\it Middle}) ; OCDM model ({\it Right}) of 
$\zs=1.0$ cases({\it Upper-panel}) and $\zs=2.0$ cases({\it Lower-panel}). 
\label{fig: varkappa}}
\end{figure}

The long-dashed lines in figure \ref{fig: varkappa} 
represent the results taking into account the finite 
mesh-size. Here, the nonlinear mass power spectrum 
$P_{\rm\scriptscriptstyle mass}(k)$ by Peacock \& Dodds (1996) 
is used in evaluating the discretized Fourier-integral 
(\ref{eq: discretize}). We also examined the various cases 
by changing the maximum number, $64<N_{\rm max}<128$, but 
obtained qualitatively similar behavior: the prediction $\derivar$ 
systematically reduces its power all over the scales, while the amplitude 
of $\var$ almost remains unchanged. Further, similar modification to 
the $z$-integral has been made so as to match the number of multiple 
lens-plane, however, this does not affect the final results.  
\clearpage

%
%
%
%
%
%
%
%
\clearpage
%
%
%
%
%




\begin{deluxetable}{lcccc}
\tablecolumns{7}
\tablewidth{8cm}
\tablecaption{Cosmological parameters used in N-body simulations. 
\label{tbl : parameter}}
\tablehead{
\colhead{Model} & \colhead{$\Omega_0$} &  \colhead{$\lambda_0$}
& \colhead{$h$} & \colhead{$\sigma_8$}
}
\startdata
~~SCDM~~~ & 1.0~~ & 0.0~~ & 0.5~~ & 0.6   \\
~~LCDM~~~ & 0.3~~ & 0.7~~ & 0.7~~ & 0.9   \\
~~OCDM~~~ & 0.3~~ & 0.0~~ & 0.7~~ & 0.85   \\
\enddata
\end{deluxetable}

\clearpage


\end{document}